\documentclass[aps,prb,jcp,floatfix,superscriptaddress,amsmath,amssymb,twocolumn]{revtex4-1}
\usepackage[utf8]{inputenc}
\usepackage[english]{babel}
\usepackage{amsmath,amssymb}
\usepackage{physics}
\usepackage{graphicx}
\usepackage[dvipsnames]{xcolor}
\usepackage{url}
\usepackage{subfigure}

\graphicspath{{plots/}}

\usepackage[colorlinks,breaklinks,linkcolor=OrangeRed,citecolor=RoyalBlue,urlcolor=NavyBlue]{hyperref}

\bibpunct{[}{]}{,}{n}{}{}

\DeclareMathOperator\arctanh{arctanh}
\newcommand{\dop}{\hat{\rho}}
\newcommand{\iu}{\mathrm{i}}
\newcommand{\Oket}[1]{|{#1}\rangle\rangle}
\newcommand{\Obra}[1]{\langle\langle{#1}|}
\newcommand{\Obraket}[2]{\langle\langle{#1}|{#2}\rangle\rangle}
\newcommand{\SF}[2]{\Pi({#1},{#2})}
\newcommand{\SFKPZ}[1]{\Pi^{\lambda}_{\text{KPZ}}\left({#1}\right)}
\newcommand{\fkpz}{f_{\text{KPZ}}}
\newcommand{\sz}{m}
\newcommand{\SzCap}{M}
\newcommand{\xxx}{{XXX }}

\newcommand{\be}{\begin{equation}}
\newcommand{\ee}{\end{equation}}

\begin{document}

\title{High-temperature spin dynamics in the Heisenberg chain: 
Magnon propagation and emerging Kardar-Parisi-Zhang scaling in the zero-magnetization limit}
\author{Felix Weiner}
\affiliation{ Institute of Theoretical Physics, University of Regensburg, D-93040 Regensburg, Germany}
\author{Peter Schmitteckert}
\affiliation{HQS Quantum Simulations GmbH, 76131 Karlsruhe, Germany}
\author{Soumya Bera}
\affiliation{Department of Physics, Indian Institute of Technology Bombay, Mumbai 400076, India}
\author{Ferdinand Evers}
\affiliation{ Institute of Theoretical Physics, University of Regensburg, D-93040 Regensburg, Germany}
\date{\today}
 
\begin{abstract}
The large-scale dynamics of quantum integrable systems is often 
dominated by ballistic modes due to the existence of stable quasi-particles.
We here consider as an archetypical example for such a system the spin-$\frac{1}{2}$ 
\xxx Heisenberg chain that features magnons and their bound states.  
An interesting question, which we here investigate numerically, arises with respect to the fate 
of ballistic modes at finite temperatures in the limit of zero magnetization $\sz{=}0$. 
At a finite magnetization density $\sz$, the spin autocorrelation function $\Pi(x,t)$ (at high temperatures)
typically exhibits a trimodal behavior with left- and right-moving quasi-particle modes 
and a broad center peak with slower dynamics. 
The broadening of the fastest propagating modes exhibits a 
sub-diffusive $t^{1/3}$ scaling at large magnetization densities, 
 $\sz {\rightarrow} \frac{1}{2}$, 
familiar from non-interacting models; it 
crosses over into a diffusive scaling $t^{1/2}$ 
upon decreasing the magnetization to 
smaller values. 
The behavior of the center peak appears to exhibit a crossover 
from transient super-diffusion to ballistic relaxation at long times. 
In the limit $\sz {\to}0$, the weight carried by the propagating peaks tends to zero; 
the residual dynamics is carried only by the central peak; 
it is sub-ballistic and characterized by a 
dynamical exponent $z$ close to the value $\frac{3}{2}$ 
familiar from Kardar-Parisi-Zhang (KPZ) scaling.
We confirm that, employing elaborate finite-time extrapolations, that the 
spatial scaling of the correlator $\Pi$ is in excellent agreement with KPZ-type behavior and analyze the corresponding corrections. 
\end{abstract}

\maketitle

%
\section{Introduction}
%
%
Interaction effects in strictly one-dimensional quantum systems tend to be strong.
This is, roughly speaking, because the dimensional reduction weakens the 
efficiency of screening and makes it difficult for 
two excitations approaching each other to avoid a collision. 
The overall reduction of phase space for (few-body) scattering 
processes has one more interesting consequence: 
Classes of one-dimensional model systems ({\em integrable}) 
can be identified that carry an extensive amount of conserved quantum numbers 
and their thermodynamic properties
can be interpreted in terms of effective particles\cite{takahashi_thermodynamics_1999}. 

When it comes to the hydrodynamic regime, it is well known that conservation 
laws tend to manifest in the analytical structures of kinetic coefficients.  
Therefore, it is an interesting endeavor to inquire into the hydrodynamics 
of fully integrable systems as has been done, recently.

Corresponding generalized hydrodynamic 
descriptions (GHD) have been 
proposed\cite{castro-alvaredo_emergent_2016,bertini_transport_2016}. 
They feature kinetic equations 
for generalized phase-space distributions, akin to the theory 
of classical soliton gases\cite{bulchandani_bethe-boltzmann_2018,doyon_soliton_2018}. 
Diffusive corrections and entropy production due to quasi-particle scattering have been 
incorporated recently\cite{de_nardis_hydrodynamic_2018,gopalakrishnan_hydrodynamics_2018,de_nardis_diffusion_2019}. 
Moreover, quantum hydrodynamics for one-dimensional systems at zero temperature developed 
earlier\footnote{The development of hydrodynamics for 1D systems is a field with a long history. We only cite one of 
the pioneering works, Ref. \cite{abanov_quantum_2005}, 
and Ref. \cite{protopopov_dynamics_2013,doyon_large-scale_2017} for a more recent 
overview},
formulated in terms of density and velocity fields, was shown to be reproduced by 
GHD in the corresponding limit\cite{doyon_large-scale_2017}.
\begin{figure}[b]
  \includegraphics[scale=0.23]{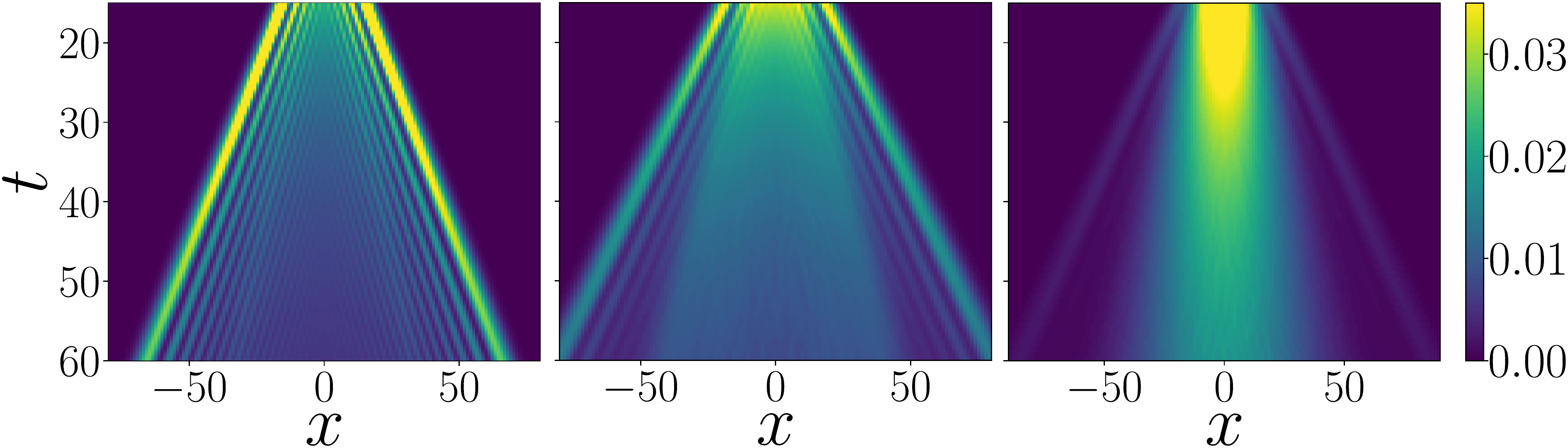}
  \caption{Spin autocorrelation function $\Pi(x,t)$ for 
  total magnetization density, $m{\approx}-0.48,-0.38,-0.17$ (left,center,right),  and $m{=}-0.17$ (right). 
  Ballistic propagation manifests itself in the ``light-cone'' structures. 
  At small magnetization a broad center peak is seen that develops a critical, 
  KPZ-type dynamics in the limit $m{\to}0$. Data has been calculated as explained in 
  the Sec. \ref{sec:method}\label{f1}\label{fig:maps}}
\end{figure}
In this work, 
we numerically investigate the spin dynamics in isotropic Heisenberg chains
at high temperatures for varying total magnetization density $\sz$. 
In the limit $\sz {\to} -\frac{1}{2}$, i.e. close to the ferromagnetic vacuum, 
only bare magnons contribute to the spin autocorrelation function $\Pi(x,t)$, 
visible as a cone in Fig. \ref{f1}.  
The front of the cone exhibits a time window of sub-diffusive 
broadening reflecting the dispersion of the quasi-particles 
(see Ref. \cite{fagotti_higher-order_2017} and references therein). 
The crossover time into the true asymptotic regime, 
which we find to be diffusive consistent with Ref. \cite{collura_analytic_2018},  
is expected to diverge in the limit $m{\to}-\frac{1}{2}$. 
Concomitantly, at intermediate $\sz$, additional ballistic modes 
can be identified corresponding to 2-magnon bound states. \\ \indent
An intriguing question arises about the 
fate of the dynamics in the limit of zero magnetization. 
In this case, it is understood that the quasi-particles 
effectively do not carry magnetization\cite{ilievski_superdiffusion_2018}. 
Consistently, we observe the weight of the propagating peaks in $\Pi(x,t)$ to disappear. 
At zero magnetization, $\sz{=}0$, and elevated temperatures, 
$T\to\infty$, a sub-ballistic dynamics of spin excitations takes over, 
as has been reported in a number of numerical studies dating back to 
\cite{fabricius_spin_1998}. However, it has long remained controversial whether a ballistic 
contribution to the spin dynamics, as measured by the Drude weight, exists (see Ref. \cite{carmelo_absence_2017} 
for an overview). Most recent analytical studies suggest that the Drude weight indeed vanishes for any finite $T>0$ 
\cite{carmelo_vanishing_2015,carmelo_absence_2017}. The peculiar residual dynamics has been 
identified as super-diffusive relaxation with a dynamical exponent close to 
$z{\approx}\frac{3}{2}$\cite{znidaric_transport_2011,ljubotina_spin_2017,richter_combining_2019,
gopalakrishnan_kinetic_2019,dupont_universal_2019}. 
Consistent with numerical observations, it was confirmed analytically 
that the spin diffusion coefficient diverges\cite{ilievski_superdiffusion_2018}. 
Recently, Ref. \cite{ljubotina_kardar-parisi-zhang_2019} showed numerically that not only the exponent $z$,
but also the spatial shape of the spin autocorrelation function is in agreement with
Kardar-Parisi-Zhang (KPZ) scaling. We confirm this result and carefully analyze 
the corrections to KPZ-scaling, which we find 
to be of the form $t^{-y}$, $y\approx 0.33$.
Our calculations for the high-temperature correlation function follow 
a protocol pioneered in Ref. \onlinecite{ljubotina_spin_2017}, employing 
standard matrix product operator (MPO) techniques. 
We observe that the magnetization dynamics with bond dimensions
$\chi \leq 1000$ exhibits unphysical fluctuations for times $t \gtrsim 30$,
which are short in comparison to the scaling limit.
Remarkably, the qualitative characteristics 
of the long-time limit, such as the dynamical exponent and the KPZ scaling, 
appears to be rather forgiving in the sense that it emerges after removing 
fluctuations by performing running averages (see Ref. \cite{ljubotina_kardar-parisi-zhang_2019}).

\section{Model and method}
\label{sec:model-method}
\subsection{Model and observable} 
The Hamiltonian of the $XXZ$ Heisenberg chain is given by
\begin{equation}
  \label{e:hamiltonianXXZ}
  \hat{H} = J \sum_{x=-\frac{L}{2}}^{\frac{L}{2}-1}\left[ \hat{S}_x^{x}\hat{S}_{x+1}^{x}+\hat{S}_{x}^{y}\hat{S}^{y}_{x+1}+\Delta \hat{S}^{z}_{x}\hat{S}_{x+1}^{z} \right]\;\text{.}
\end{equation}
We choose $J{=}1$ as the unit of energy and concentrate on $\Delta {=}  1$\footnote{The sign of $\Delta$ is insignificant for the observables studied in this work}. The total $z$-component of spin $\hat{M} = \sum_x \hat{S}^z_x$ is conserved, i.e. commutes with $\hat{H}$. The length of the chain is chosen $L{\geq}200$ such that, on the time scales shown, the boundaries do not affect the results of this work. 

Our observable is the spin dynamics by means of the equilibrium $S^z$ correlation function:
\begin{align}
  \label{e:pi-definition}
  \SF{x}{t} = \mathcal{N}\left( \langle \hat{S}^{z}_x(t)\hat{S}^{z}_0 \rangle_h - \langle \hat{S}^{z}_0 \rangle_h^2  \right)
\end{align}
with $\hat{S}^z_x(t)=e^{\iu \hat{H} t}\hat{S}_x^z e^{-\iu\hat{H}t}$. Averages are taken with respect to an infinite temperature ensemble
\begin{equation}
  \label{e:ensemble}
  \langle \hat{X} \rangle_h = \frac{\Tr\left[e^{-h\hat{M}} \hat{X} \right]}{\Tr e^{-h\hat{M}}} \, \text{,}
\end{equation}
where $h$ controls the average magnetization $\langle \hat{M} \rangle_h = \frac{L}{2} \tanh\left( \frac{h}{2} \right)$. The prefactor in \eqref{e:pi-definition} is time-independent and normalizes the correlator: $\displaystyle \sum_{x} \SF{x}{t} = 1$.

Instead of directly evaluating $\SF{x}{t}$, we adopt the simulation protocol suggested in \cite{ljubotina_spin_2017} and 
compute the time evolution of a non-equilibrium initial state  
\begin{equation}
\label{e:init_state}
  \dop_0 \sim \exp\left(-h\hat{M} - \displaystyle\sum_{x} \mu_x\hat{S}^z_{x}\right) 
\end{equation}
corresponding to a high temperature state with varying $\SzCap$-density. The initial spin profile has a
``domain wall''-shape: 
\begin{equation}
  \label{e:dom_wall}
\mu_x = \left\{
\begin{array}{ll}
+\mu, & x > 0 \\
-\mu, & x \leq 0 \\
\end{array}
\right.
\end{equation}
Then, in the limit of small $\mu$, the non-equilibrium spin densities can be related to the 
equilibrium correlator $\SF{x}{t}$: 
\begin{equation}
\label{e:relation_w_SF}
\partial_x \Tr\left[\rho_0 \hat{S}^z_x(t)\right] \approx \mu\left( \langle S^{z}_x(t)S^{z}_0 \rangle_h - \langle S^{z}_0 \rangle_h^2  \right)  + \mathcal{O}(\mu^2)\,\text{.}
\end{equation}
The spatial derivative is evaluated numerically. While such a linear-response relation is easily seen to hold for continuous $x$, an exact relation for the lattice model was derived in Ref. \cite{ljubotina_kardar-parisi-zhang_2019}. We chose $\mu = 0.001$ in the numerical simulations.
\subsection{Method\label{sec:method}} 

Time evolution $\dop(t) = e^{-\iu \hat{H} t}\dop(t{=}0)e^{\iu \hat{H} t}$ is carried out using a matrix product decomposition of \eqref{e:init_state} (controlled by the maximum bond dimension $\chi$) and  conventional Trotter decomposition (controlled by the time increment $\Delta t$) of the Liouvillian superoperator $\mathcal{L}$ corresponding to the Hamiltonian \eqref{e:hamiltonianXXZ}: $\mathcal{L}\dop \equiv [\hat{H},\rho]$. Conservation of $\hat{S}^z$ implies a block-diagonal structure of $\dop$, which is exploited in order to speed up the calculations. As these are standard techniques used in the field, further details are delegated to the Appendix \ref{sec:matrix-product-state}. The calculations were performed using the ITensor library\cite{stoudenmire_itensor_nodate}. 

We offer a remark concerning the convergence with bond dimenion, $\chi$. 
Quite generally, the convergence properties with respect to $\chi$ are far from universal: 
depending on the system (i.e. the Hamiltonian), the initial state and certain computational details, 
convergence of a given observable can be reached at significantly different $\chi$ values. 
Indeed, a remarkable observation was made in \cite{ljubotina_spin_2017}: 
Evolving initial states of type \eqref{e:init_state}, the authors could obtain results 
for the spin and current densities that are roughly independent on $\chi$ up to long times $t \lesssim 150$. 
This exceeds simulation times reported for direct evaluation of correlation functions by almost an order of magnitude (see e.g. Refs. \cite{barthel_precise_2013,kennes_extending_2016} for an analysis of convergence properties). 

While our results fully confirm the qualitative conclusions of \cite{ljubotina_spin_2017,ljubotina_kardar-parisi-zhang_2019}, we do observe corrections upon increasing $\chi$. For example, the diffusion constant at $\Delta{=}2$ is observed to keep increasing with $\chi > 1000$ where its value has increased to $D{\geq}0.63$ as opposed to 
$D \approx 0.4$ reported in \cite{ljubotina_spin_2017} (see Appendix \ref{sec:diff-const-at}). 
Based on the impression that results may still exhibit a significant dependence on the bond dimension, the convergence of dynamical properties with $\chi$ will receive a special attention below. An additional discussion of convergence behavior is given in Appendix \ref{sec:convergence}. 

\begin{figure}
  \centering
  \includegraphics[scale=0.57]{{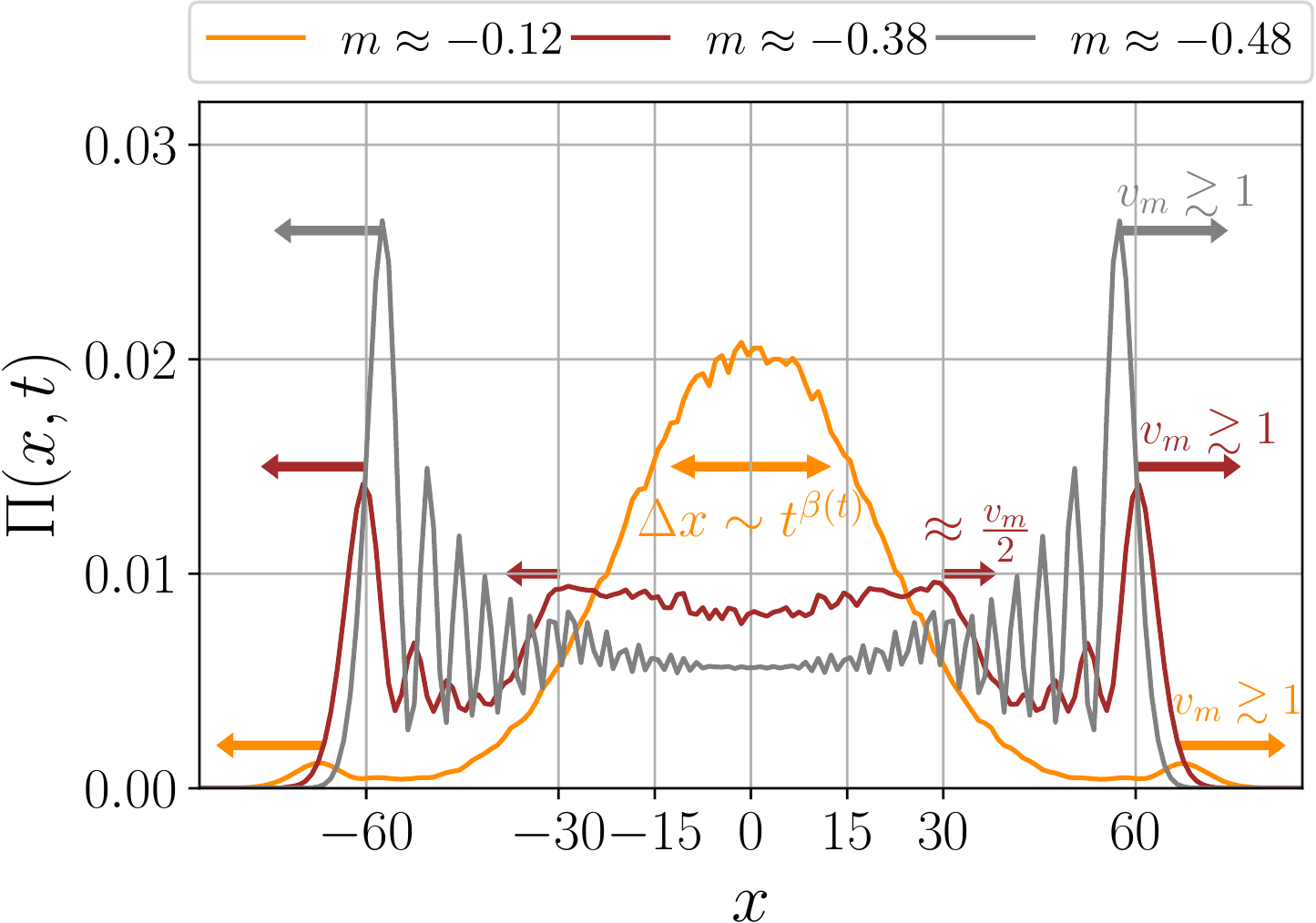}}
  \caption{Spin density profiles at time $t{=}60$ for varying magnetization density $\sz$. The outer peaks at $|x| {\approx} v_{\sz}t {\approx} 60$ correspond to propagating magnon-like excitations with a velocity $v_{\sz} {\gtrsim} 1$, which is slightly renormalized with increasing $m$. For $\sz{=}-0.38$, distinct peaks can be observed around $x{\approx}v_{\sz} t/2 {\approx} 30$, which we attribute to 2-magnon bound states. The center peak seen for $m{=}-0.12$ exhibits sub-ballistic broadening $\sim t^{\beta_{\sz}(t)}$ on the time scales shown here. In the limit of $\sz\rightarrow 0$, our results are consistent with an exponent of $\beta_{\sz=0}(t\rightarrow \infty) = \frac{2}{3}$ as observed in previous studies.  
  \label{fig:var_mg}}
\end{figure}

\section{Results\label{sec:results}}

The form of the correlation function \eqref{e:pi-definition} is determined by the motion of the quasi-particles. We refer to Ref. \cite{gopalakrishnan_anomalous_2019} for a discussion of the analytic properties of this correlator.  In the limit of $\sz = \langle \hat{M}\rangle_h/L \rightarrow -0.5$, the correlator $\Pi$ probes dynamics close to the fully polarized state. Excitations of this state are magnons and bound states of $n$ magnons with bare group velocity $v_b \sim \frac{J}{n}$\cite{takahashi_thermodynamics_1999}. 
Due to the integrability of the model, quasi-particles remain stable even for $|\sz| < 0.5$ and give rise to ballistic modes in the spin dynamics, which are observed in $\Pi(x,t)$ as a set of propagating peaks. The spatial dependence of $\Pi$ at fixed time for varying $\sz$ is illustrated in Fig. \ref{fig:var_mg}: At strong magnetization $|\sz|{\to}0.5$ only magnons contribute to $\Pi$, which manifest themselves as a sequence of left- and a right-moving peaks with velocities $\pm v_{\sz}$. 
The evolution of these peaks and their dependence on $\sz$ is analyzed in 
Sec. \ref{sec:finite-sz}. At intermediate $|\sz|$ we can also identify another 
pair of distinct peaks in $\Pi$, which move with a slower velocity that is 
given by roughly half the magnon velocity (as indicated by the arrows in the figure). 
Therefore, they can be associated with 2-magnon bound states. 

Upon further decreasing $|\sz|$, only a single propagating peak in $\Pi$ is left,
the remaining weight is carried by a broad peak centered around $x=0$, 
at least on the time scales studied here. This peak exhibits anomalous KPZ scaling at 
vanishing total magnetization $\sz{=}0$, which we analyze in Sec. \ref{sec:comp-kpz}. 
The behavior of the center peak for finite $\sz$ is discussed in Sec. \ref{sec:fate-kpz-peak}.
\begin{figure*}
  \centering
  \mbox{    
  \subfigure[\label{fig:front_largem}]{\includegraphics[scale=0.395]{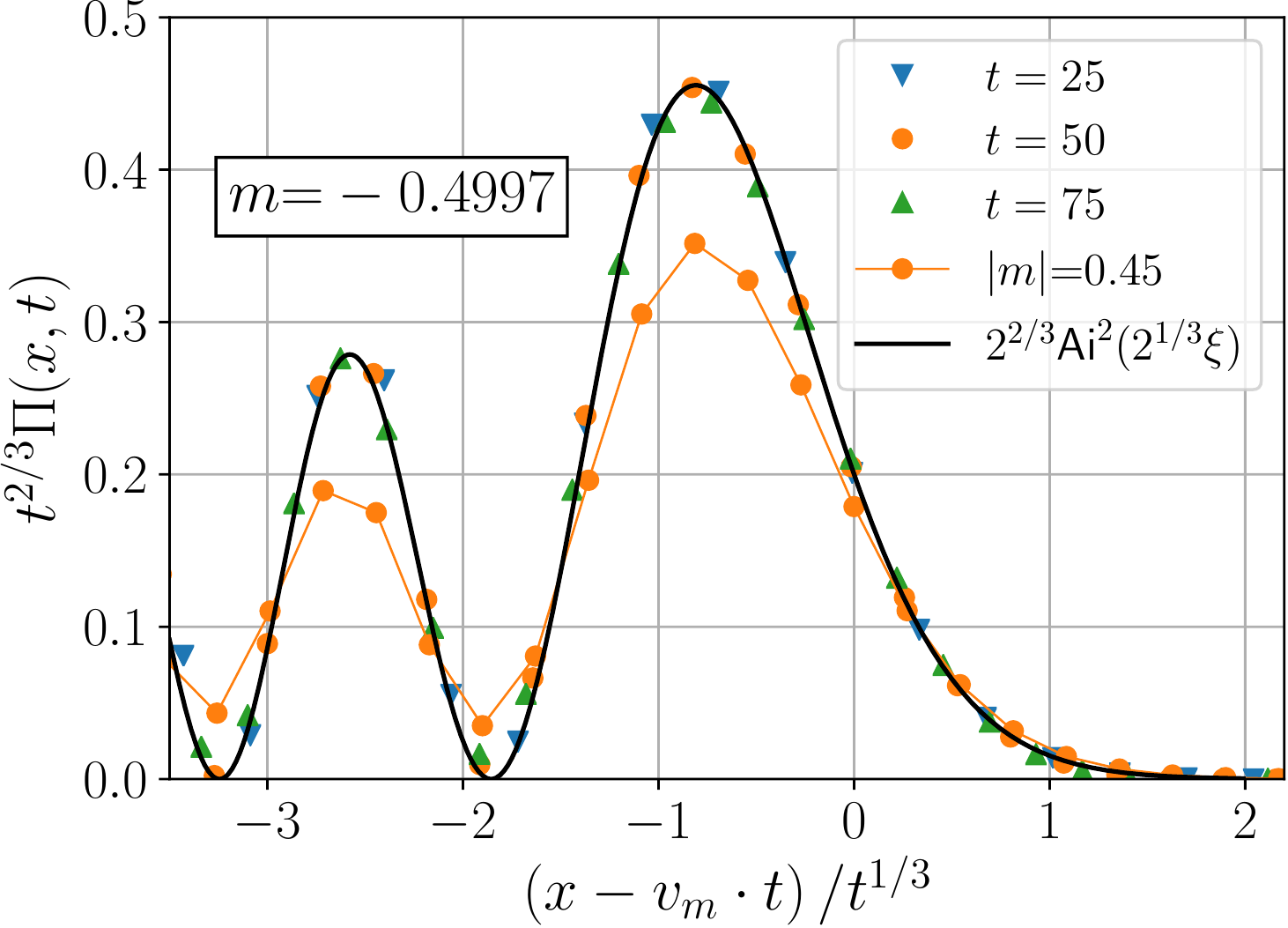}}
  \subfigure[\label{fig:front_intermediatem}]{\includegraphics[scale=0.395]{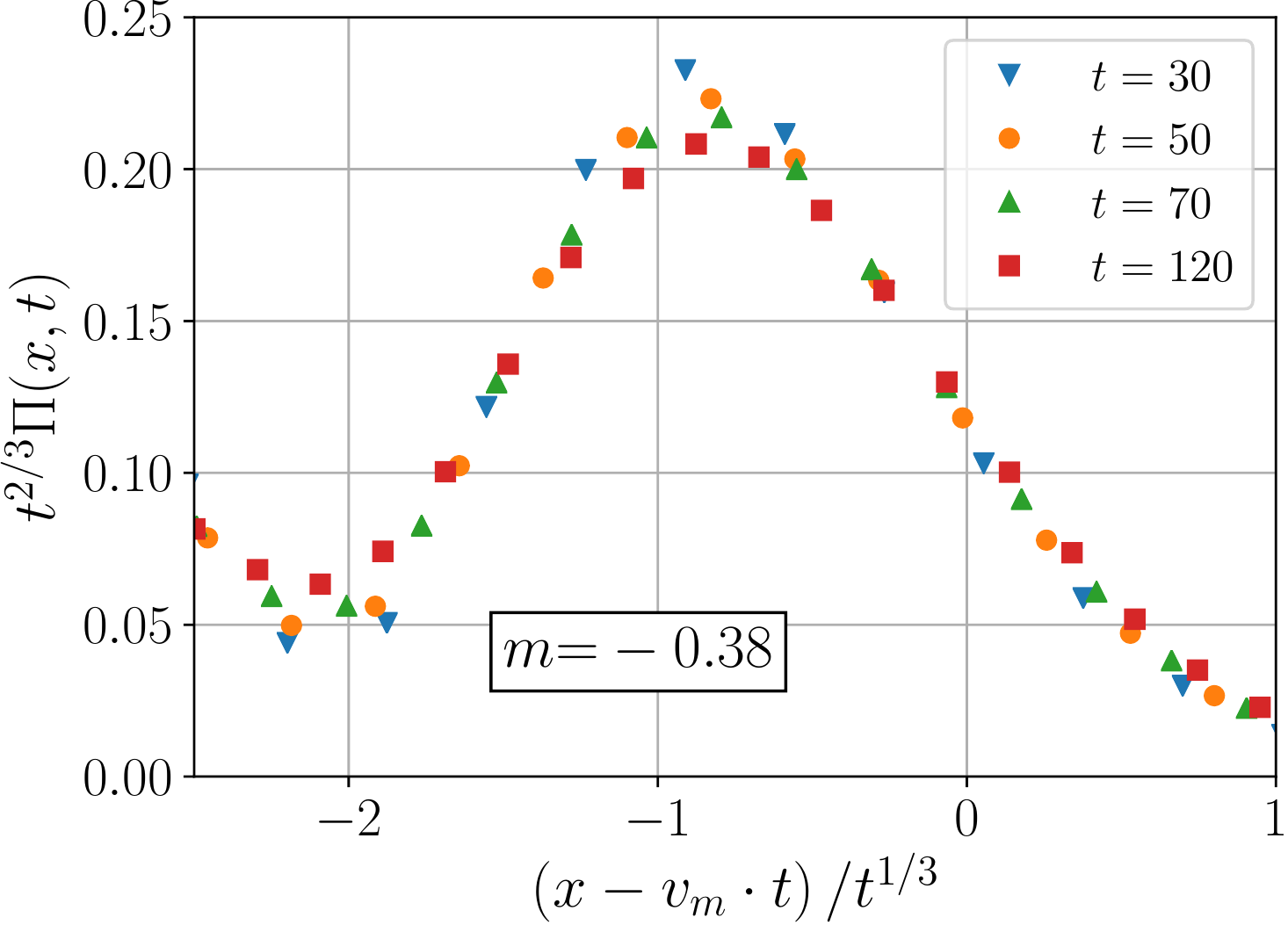}}
  \subfigure[\label{fig:front_smallm}]{\includegraphics[scale=0.395]{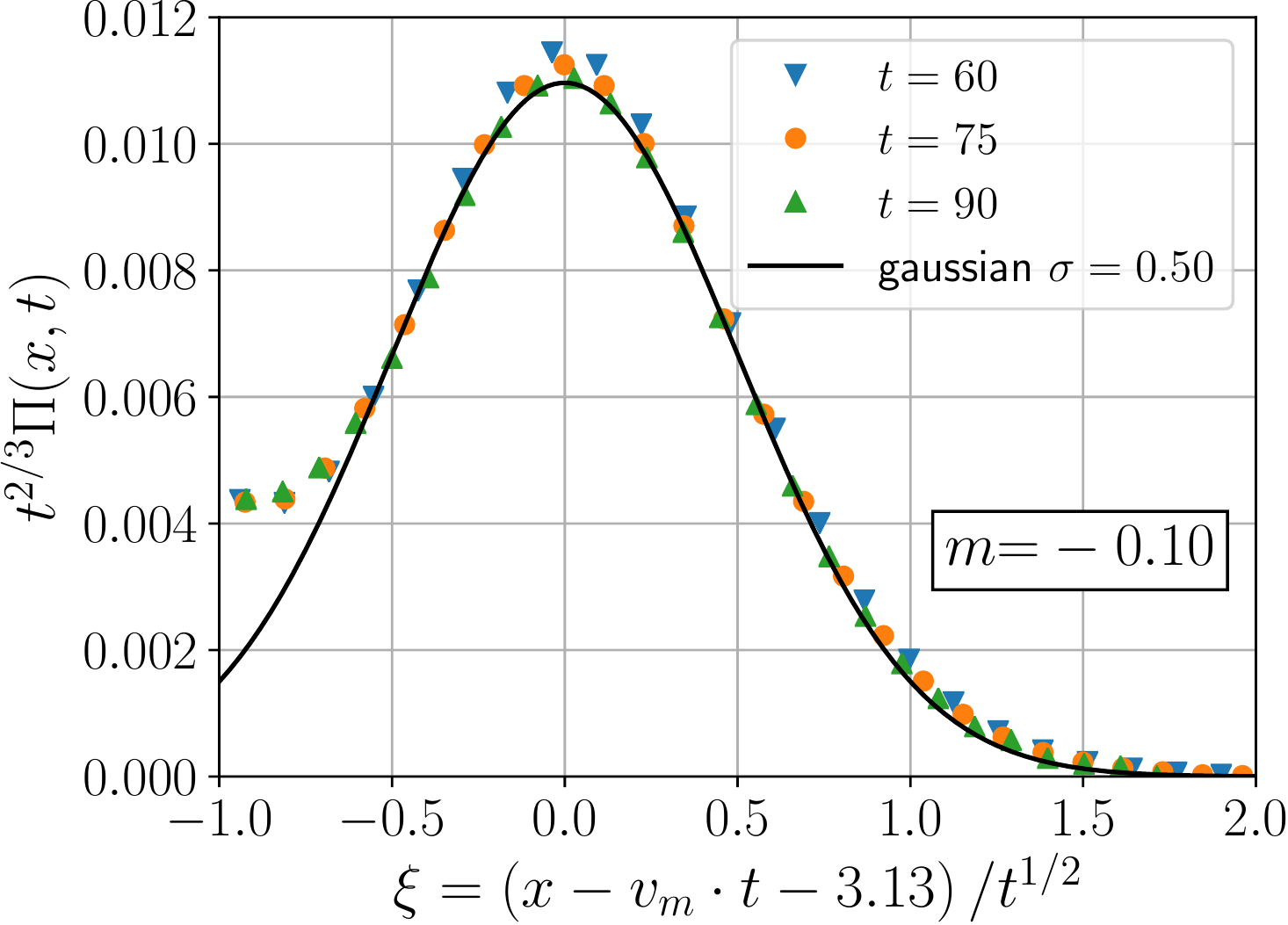}}
}
\caption{Rescaled correlator close to the light-cone $x{=}v_{\sz} t$ for different values of $\sz$. The figures illustrate the evolution of the magnon peak as a function of magnetization $m$ and time $t$. \textbf{(a)} In the strongly magnetized case $\sz \approx -0.5$, numerical results are compared with the exact scaling function for non-interacting systems\cite{hunyadi_dynamic_2004}. We also show data for a slightly smaller value of $|m|$, which illustrates how the profile is smeared out away from the fully polarized limit. \textbf{(b)} For an intermediate value of the magnetization, $|m|{\approx}0.38$, we find that the scaling collapse, as observed in (a), holds only approximately at relatively short times. In order to illustrate the deviations, we also show data for $t=120$. \textbf{(c)} At $\sz \approx -0.1$, the numerical data can be described by a Gaussian close to the peak, i.e. around $\xi{=}0$. Note the different rescaling of the spatial coordinate as compared to Figs. \ref{fig:front_largem} and \ref{fig:front_intermediatem}. In the weakly magnetized limit, we found that a diffusive rescaling of the spatial coordinate yields a better collapse of the numerical data (see discussion in the main text).
\label{fig:front}}
\end{figure*} 

\subsection{Finite magnetization: Magnon modes\label{sec:finite-sz}}
\subsubsection{Shape of the magnon peak\label{sec:magnon-fronts}}

In the limit $\sz {\rightarrow} -0.5$, the correlator $\Pi(x,t)$ probes spin dynamics close to the fully polarized state, which is the (grand-canonical) ground state of the ferromagnetic Heisenberg chain. The fastest excitations of this state are free magnons with bare dispersion $\epsilon(k) {\sim} -J \cos(k)$ and a maximum group velocity of $v{=}J({=}1)$. 

In the context of non-interacting models, it is well known that non-linearity of the free quasi-particle dispersion gives rise to a peculiar scale-invariance of density profiles close to the ``light-cone''\cite{hunyadi_dynamic_2004,eisler_full_2013}, i.e. for $x{=}{\pm}vt$ (where $v$ denotes the velocity of the fastest quasi-particle mode). More specifically, the broadening of the ballistic front is given by a sub-diffusive power law $t^{1/3}$. Recently, attempts have been made to interpret these findings in the context 
of GHD \cite{fagotti_higher-order_2017,bulchandani_subdiffusive_2019}. We would like to stress that the exponents arising here, while similar to the KPZ exponents, 
are believed to have a different origin as the dynamical exponent $z$ discussed in Sec. \ref{sec:superd-expon} below. The latter is interpreted as a consequence of interactions between quasi-particles.

In Fig. \ref{fig:front_largem} we show that, for large magnetization ($m \approx -0.5$), 
$\Pi$ appears to exhibit the sub-diffusive scaling close to the light-cone, as in the non-interacting case:
\begin{equation}
  \label{e:scaling_bessel}
  \Pi(x,t) \sim \frac{1}{t^{2/3}}F\left(\frac{x\pm v_{\sz}t}{t^{1/3}}\right)
\end{equation}
for $x {\sim} v_{\sz}t$, where the $F(y){=}2^{2/3}\text{Ai}^2(2^{1/3}y)$ with $\text{Ai}(y)$ denoting the Airy function\cite{hunyadi_dynamic_2004}. In order to achieve a collapse of our numerical data for different times, we need to account for a small renormalization of the bare magnon velocity (the actual velocity $v_{\sz}$ is taken as a fit parameter here). In fact, fitting the profiles to the function $F(y)$ can be used as a method to extract the velocities as long as $h \gtrsim 1$, as discussed below. 

The question of whether the $t^{1/3}$ scaling survives in the presence of interactions has been discussed in recent 
works \cite{fagotti_higher-order_2017,collura_analytic_2018,bulchandani_subdiffusive_2019}. In Ref. 
\cite{bulchandani_subdiffusive_2019}, it was shown numerically to occur for any values of $\Delta$ if the initial state 
is given by a polarized product state, consistent with our findings for the spin correlator. On the other hand
for more generic non-equilibrium situations, it is 
expected that diffusive dynamics\cite{de_nardis_diffusion_2019} will 
eventually dominate over the dispersive $t^{1/3}$ scaling in the long-time limit. 
Numerical evidence for such a diffusive scaling of spin profiles close to the light-cone was given in Ref. \cite{collura_analytic_2018}
in the regime $\Delta < 1$. In the following, we study the crossover in more detail, as a function of both magnetization and time.

\begin{figure}
  \centering
  \includegraphics[scale=0.42]{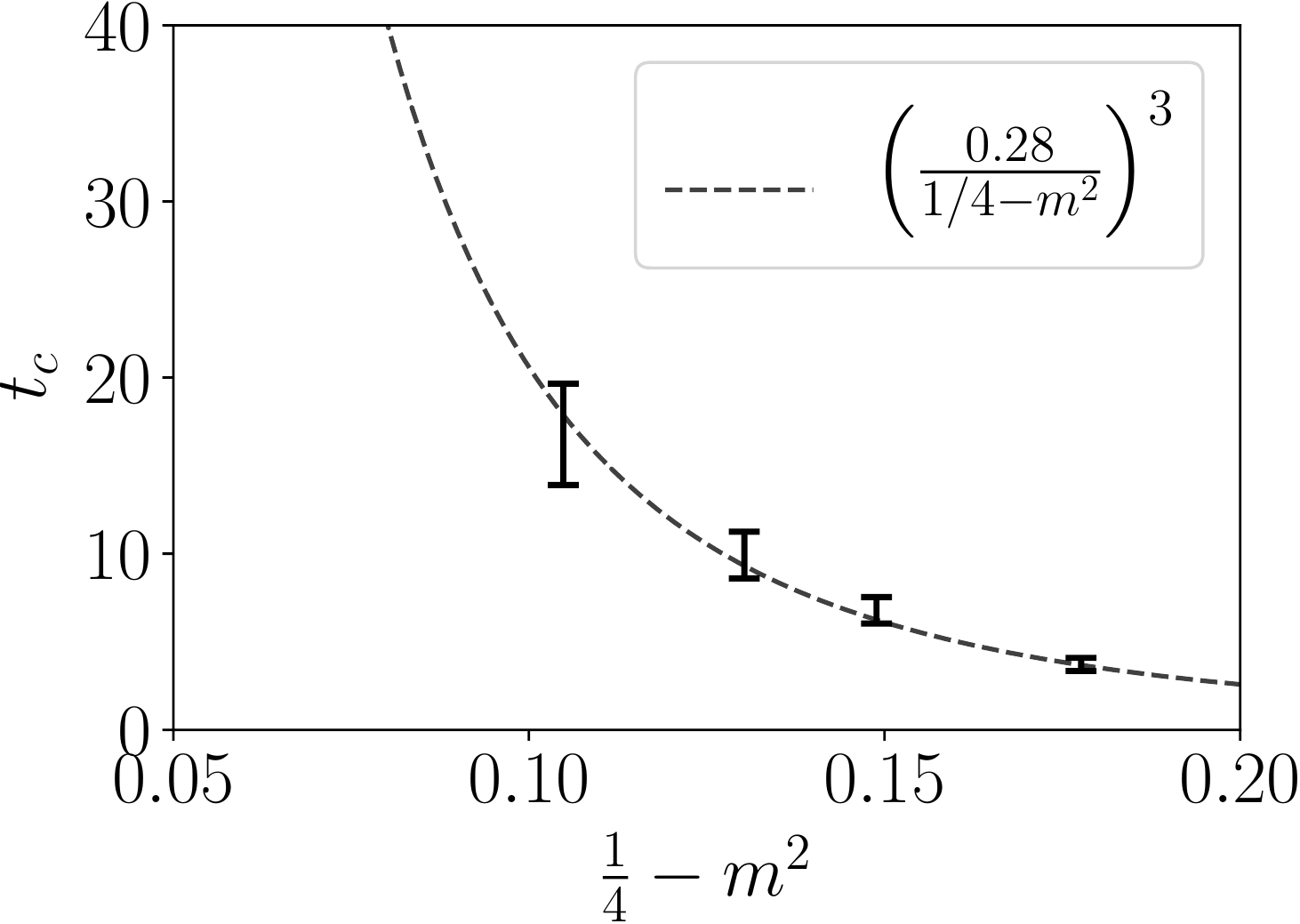}
  \caption{Estimate for the crossover time scale that separates the diffusive ($t> t_c$) from the sub-diffusive ($t<t_c$) growth of the width of the fastest propagating peak. The crosses indicate numerical results, while the dashed line is a conjecture based on Eq. \eqref{eq:e:dc_magnon}. Error bars reflect the residual time dependence of the numerical values obtained for $D_{\text{mag}}$.\label{fig:crossover}}
\end{figure}

Upon decreasing $|\sz|$, going away from the fully polarized limit, we observe that the features of $F(y)$, as given in Eq. \eqref{e:scaling_bessel}, are increasingly washed out. We indeed find that, for small enough $|m|$, the broadening of the magnon peak appears to follow a diffusive $t^{1/2}$ scaling at variance with the sub-diffusive $t^{1/3}$ scaling observed for $|m|{\to}0.5$ (see Fig. \ref{fig:front_smallm}). 
From the reasoning above, one would expect that the width of the fastest propagating peak, $\sigma_{\text{mag}}$, can be described by $\sigma^2_{\text{mag}}(t) = D_{\text{mag}}t + (\kappa t)^{2/3}$. The latter prefactor, $\kappa$, can be determined from our data in the limit $|m|{\rightarrow}\frac{1}{2}$, see Fig. \ref{fig:front_largem}, which yields $\kappa \approx 0.18$. Stipulating that $\kappa$ is independent of the magnetization $m$, we define a crossover time scale
\begin{equation}
  \label{e:crossover}
  t_c \sim \frac{\kappa^2}{D_{\text{mag}}^3} \approx \frac{0.03}{D_M^3}\,\text{,}
\end{equation}
such that for $t \ll t_c$ the broadening is sub-diffusive, $\sigma_{\text{mag}} \sim t^{1/3}$, and for $t \gg t_c$ it is diffusive $\sigma_{\text{mag}} \sim t^{1/2}$. For small and intermediate values of $|m|$, estimates for the prefactor $D_{\text{mag}}$ can be obtained from numerical data, e.g. from Gaussian fits as seen in Fig. \ref{fig:front_smallm}. In this manner, we obtain the estimates for $t_c$ shown in Fig. \ref{fig:crossover}. In the strongly magnetized limit, the diffusion constant associated with the broadening, $D_{\text{mag}}$, is expected to be determined by the magnon occupation factor $\theta_1$ only\cite{gopalakrishnan_kinetic_2019,ilievski_superdiffusion_2018}:
\begin{equation}
  \label{eq:e:dc_magnon}
  D_{\text{mag}}\sim \theta_1(1-\theta_1) \overset{|m|\to 1/2}{\approx} \theta_1 = (1/4 - m^2)\,\text{,}
\end{equation}
while contributions from bound states are suppressed. Indeed, we find that our numerical results for $t_c$ are consistent with a divergence of the form $\sim \left( 1/4 - m^2\right)^{-3}$ upon approaching the fully magnetized limit $|m|=\frac{1}{2}$ (see Fig. \ref{fig:crossover}). 

\begin{figure}
  \centering
  \includegraphics[scale=0.42]{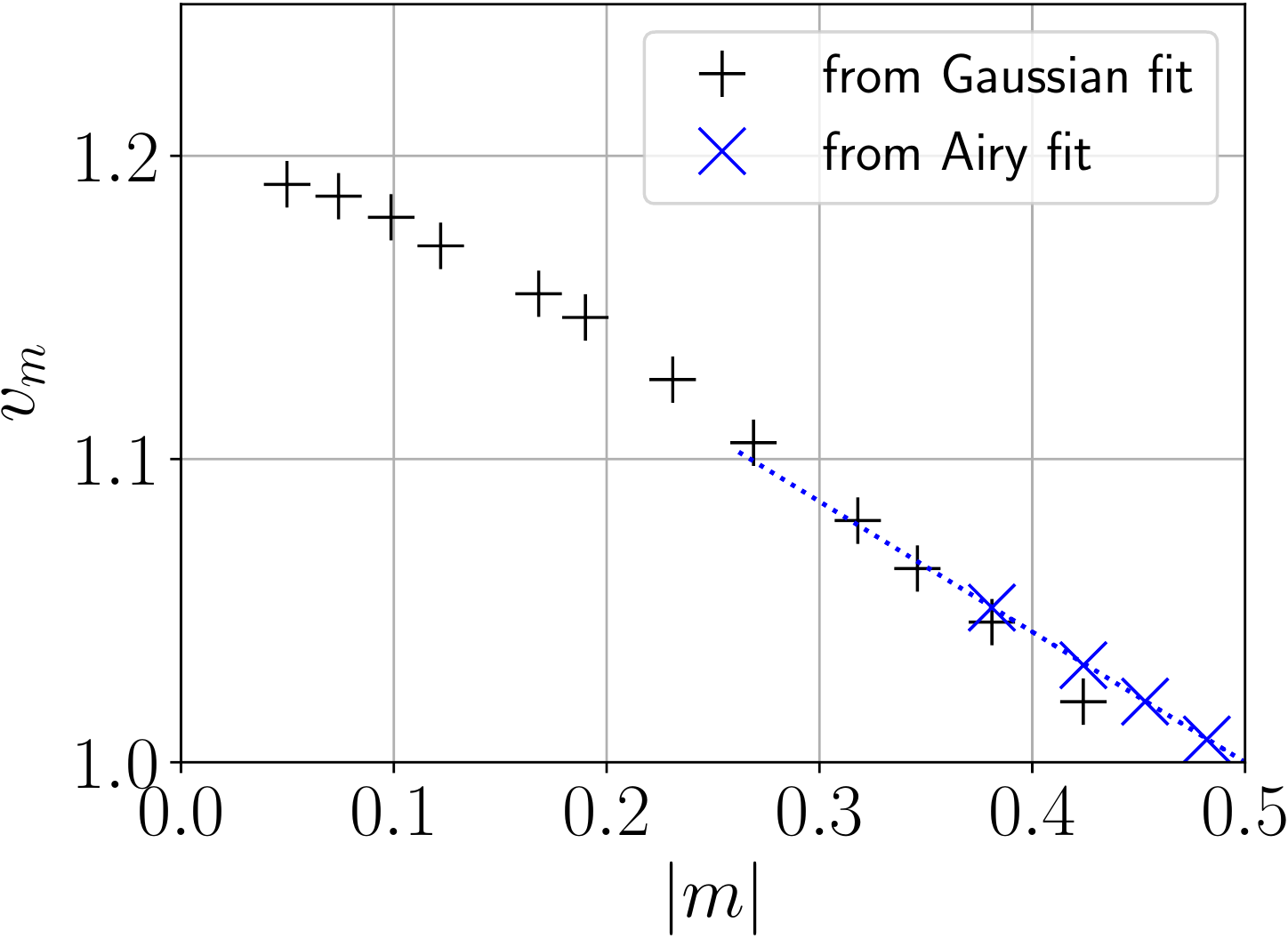}
  \caption{Renormalized velocities of the ``magnon peak'' as a function of the magnetization density $m$. The dashed line corresponds to a linear fit of the blue data points, which have been obtained obtained from fitting the ballistic peaks to the scaling function \eqref{e:scaling_bessel} (cf. Fig. \ref{fig:front_largem}).\label{fig:vel}}
\end{figure}

\begin{figure*}[th]
  \centering
  \subfigure[\label{fig:dx}]{\includegraphics[scale=0.395]{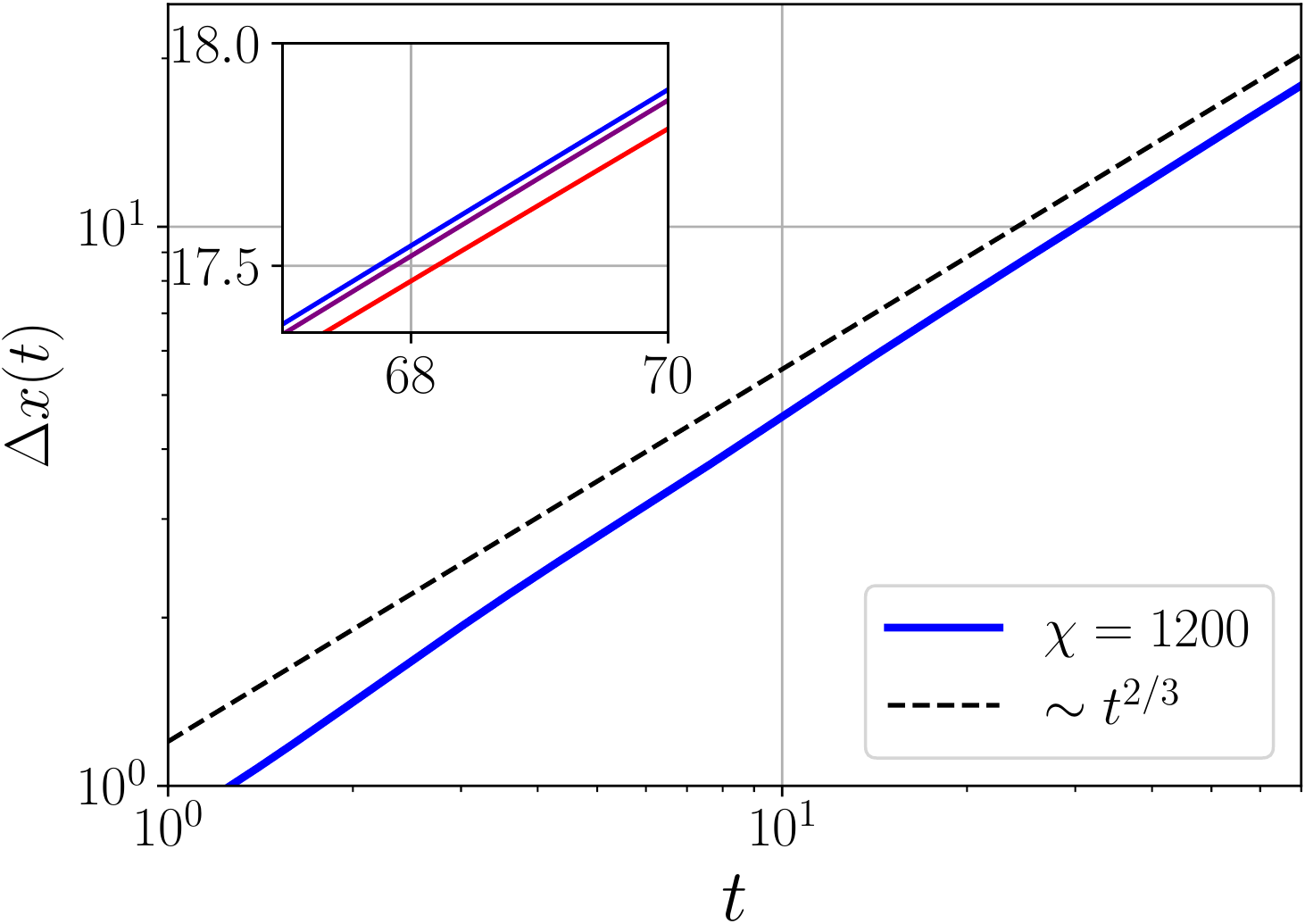}}
  \subfigure[\label{fig:beta}]{\includegraphics[scale=0.395]{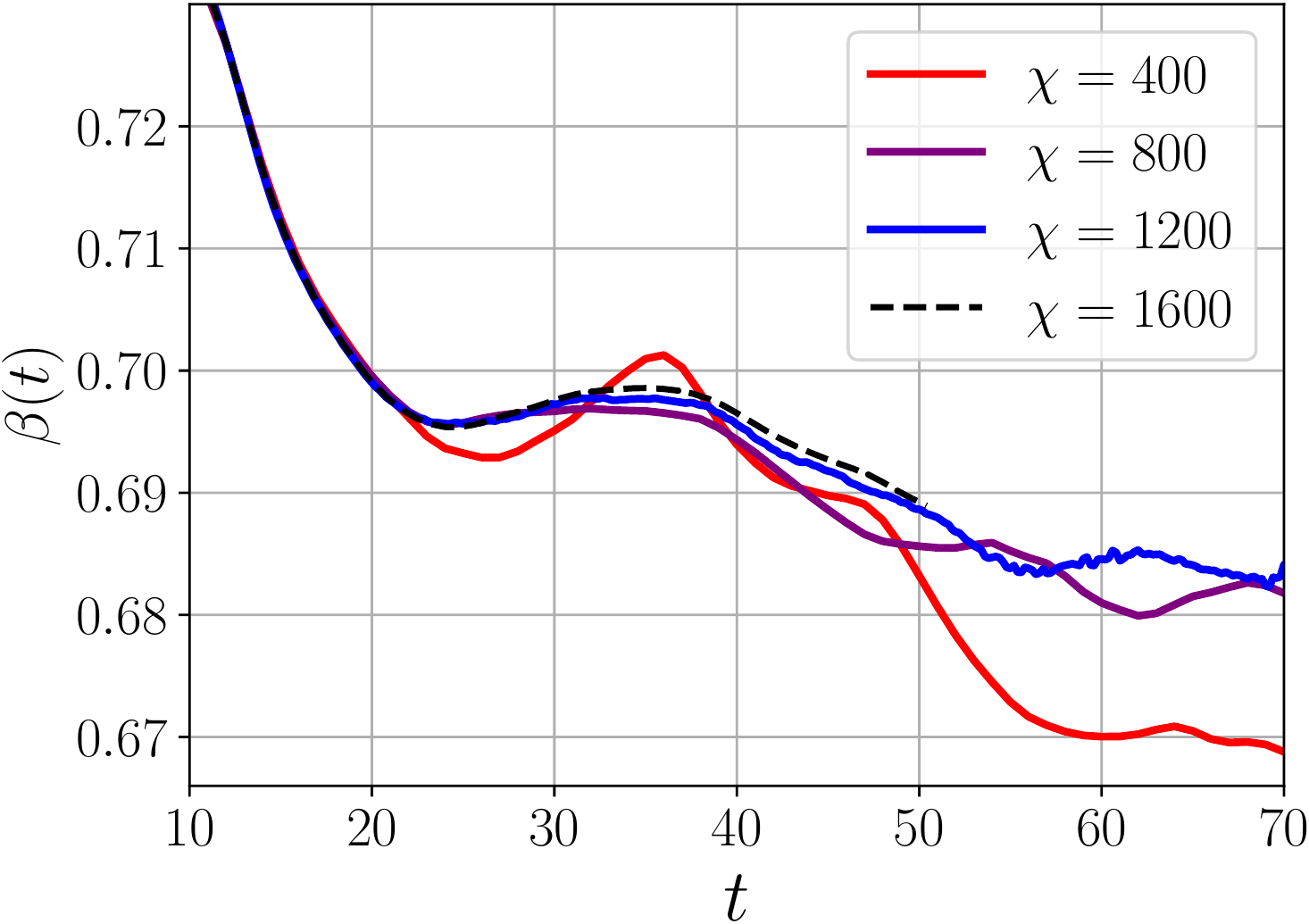}}
  \subfigure[\label{fig:dx_corr}]{\includegraphics[scale=0.395]{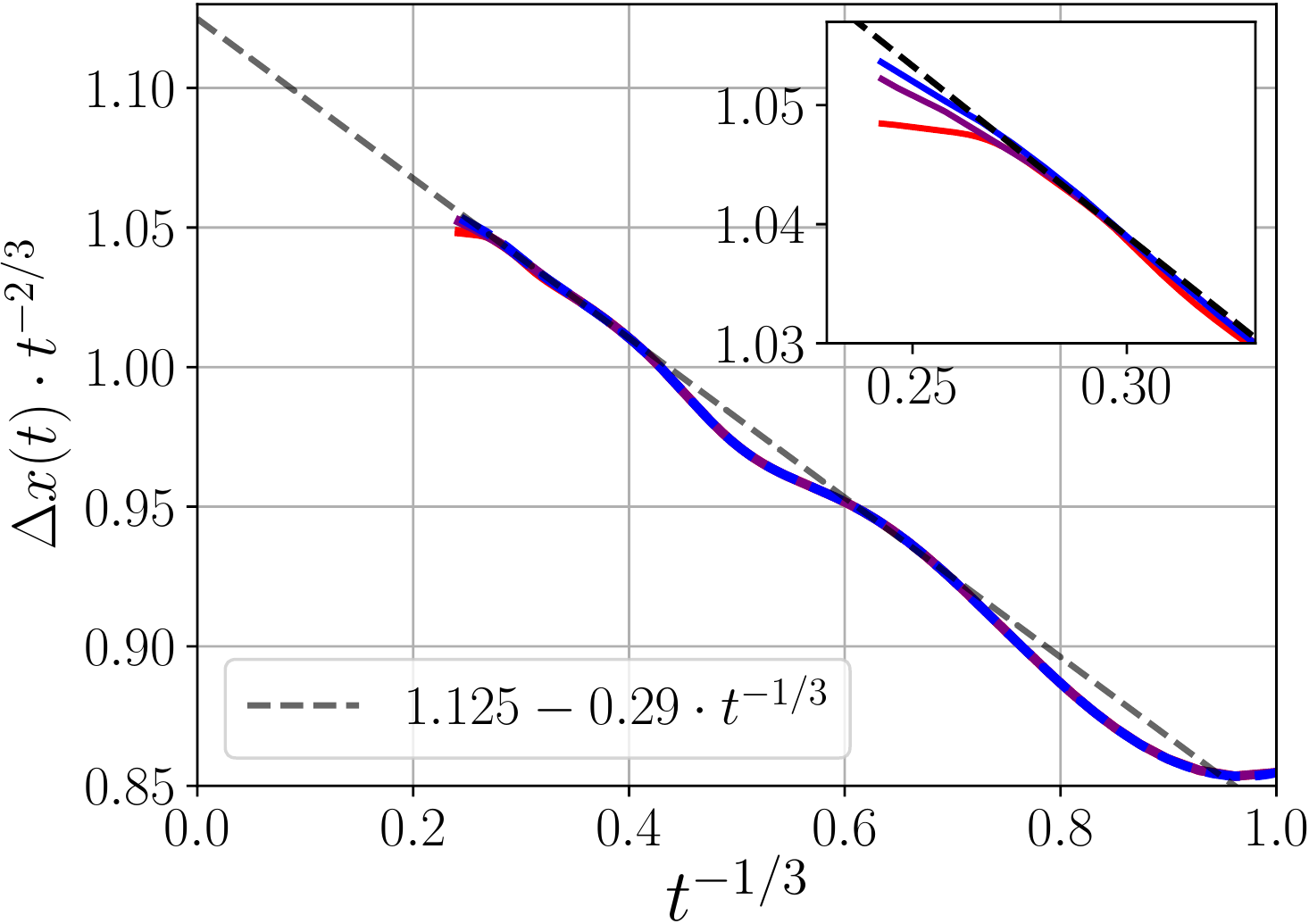}}
  \caption{
  Characteristics of the spin dynamics computed from $\SF{x}{t}$
  at zero magnetization, $\sz{=}0$. 
  \textbf{(a)} Time dependence of the width $\Delta x$ of 
the correlator. The black line serves as a guide to the eye, 
indicating a power law $t^{1/z}$ corresponding to the KPZ 
exponent $z=\frac{3}{2}$. The inset highlights the dependence 
on the maximum bond dimension $\chi$, which is not visible 
in the main plot. \textbf{(b)} The effective, time dependent exponent 
$\beta(t)$ (see Eq. \eqref{e:eff-exp}) 
highlighting deviations from true power law behavior  and significant residual dependency on the bond dimension $\chi$. 
\textbf{(c)} The graph suggests that corrections to the power law behavior can be described by a subleading term $\sim t^{1/3}$. The offset and slope obtained from the fit (dashed black line) are discussed in the main text.\label{fig:dx_analysis}}
\end{figure*}

\subsubsection{Velocity\label{sec:velcity}}

Tracking the outermost peak appears to be the simplest scheme for extracting renormalized magnon velocities. We achieve this by fitting the numerical data to a Gaussian close to the peak and then obtain the velocity via linear regression. The results are shown in Fig. \ref{fig:vel}. For large magnetization $|m|$, however, such a scheme does not yield accurate results on the time scales $t\leq 100$ studied in this work. The reason for this failure becomes obvious from the scaling shown in Fig. \ref{fig:front_largem}: $\xi{=}0$ does not correspond to the position of the peak but rather to a different point at larger $\xi$ (corresponding to the turning point of the Airy-function). Therefore, the position of the peak $x_p(t)$ exhibits a subleading term $x_p=v_{\sz}t + \text{const}\cdot t^{-2/3}+ \ldots $ at short times, when the sub-diffusive scaling still holds approximately. We also show velocities obtained from fitting the profile to the function $F(y)$ ( see 
Eq. \eqref{e:scaling_bessel} ) at large $|m|$. The thus obtained values linearly extrapolate to the correct bare magnon velocity: $v_{\sz} \approx 1.0 + 0.43 \cdot (|\sz| - \frac{1}{2})$ for $\sz\gtrsim -\frac{1}{2}$.
In the opposite limit, $m{\rightarrow}0$, the renormalized magnon velocity appears to approach a value of $v_{\sz=0}\approx 1.2$, which is consistent with the value for the ``Lieb-Robinson'' velocity at vanishing magnetization that was obtained in \cite{ilievski_superdiffusion_2018} (see Fig. 1, inset, of that reference). 

\subsection{Spin profiles at $\sz = 0$: KPZ scaling \label{sec:comp-kpz}}

A recent numerical work \cite{ljubotina_kardar-parisi-zhang_2019} studied the high-temperature spin correlator \eqref{e:pi-definition} in the isotropic Heisenberg chain at vanishing total magnetization, i.e. $\sz{=}0$. Interestingly, the authors found that the spatial profile is given by scaling functions of the KPZ universality class, consistent with the dynamical exponent $z{=}\frac{3}{2}$ observed earlier\cite{znidaric_transport_2011,ljubotina_spin_2017}. In this section we confirm these observations by carefully analyzing transients  and corrections to scaling, 
as well as the dependence of the numerical results on the bond dimension.    

The KPZ-equation was initially suggested as a description of universal properties of surface 
growth\cite{kardar_dynamic_1986}. The closely related stochastic Burger's equation appears as a hydrodynamic limit in 
many classical many-body systems in one dimension (see eg. Ref. \cite{van_beijeren_exact_2012}). Manifestations of KPZ 
universality in quantum systems are subject to on-going research (see Refs. 
\cite{kulkarni_finite-temperature_2013,arzamasovs_kinetics_2014,poboiko_spin_2016,samanta_thermal_2019} for works 
outside of the present context). It should be noted, however, that a theoretical understanding of why KPZ universality 
emerges in the integrable \xxx chain is still lacking. Some aspects of the super-diffusive dynamics have been captured 
by a kinetic theory\cite{gopalakrishnan_kinetic_2019}. Furthermore, numerical studies have provided insight regarding 
the relevant conservation laws: A recent study indicates that integrability is indeed a crucial ingredient in order to 
observe a dynamical 
exponent 
$z=\frac{3}{2}$ in spin chains\footnote{See Ref. \cite{dupont_universal_2019}. It should be noted, however, that previous studies found super-diffusive dynamics also in the 
non-integrable spin-$1$ Heisenberg chain\cite{de_nardis_anomalous_2019,richter_magnetization_2019} at high temperatures.}; 
the relevancy of energy conservation is presently investigated\cite{ljubotina_kardar-parisi-zhang_2019}. 

\subsubsection{Time evolution of $\Delta x(t)$\label{sec:superd-expon}}
As ballistic contributions are absent at $\sz {=} 0$, the dynamics of the center peak is characterized by the width
\begin{equation}
  \label{e:rmsd}
\Delta x(t)= \left( \sum_x x^2 \SF{x}{t} \right)^{1/2}\text{.}
\end{equation}
This quantity can be interpreted as the root-mean-squared displacement of an excess spin density initially localized at the origin $x{=}0$. The corresponding numerical data is displayed in Fig. \ref{fig:dx}, exhibiting an approximate power law $t^{1/z}$ with dynamical exponent $z\approx 1.5$. $z$ being close to $\frac{3}{2}$ has been observed before\cite{znidaric_transport_2011,ljubotina_spin_2017} and was giving a motivation to inquire into the possibility of KPZ dynamics. 

\paragraph*{Convergence of effective exponent function} 
In order to highlight the deviations from a true power law behavior 
as well as the dependence on the bond dimension $\chi$, we introduce the effective exponent
\begin{equation}
  \label{e:eff-exp}
  \beta(t) = \frac{d\,\log \Delta x(t)}{d\,\log(t)}\text{.}
\end{equation}
Results are shown in Fig. \ref{fig:beta}. While saturation of $\beta(t)$ near a value of $\frac{3}{2}$ is 
observed at relatively small $\chi$, deviations grow at better $\chi$-values; 
concomitantly, the ``noise'' on $\beta(t)$ seen in Fig. \ref{fig:beta} diminishes. 
Strictly speaking,  the asymptotic value 
$\beta(t{\rightarrow}\infty)$ is not reliably obtained from the data 
without further analysis. 

\paragraph*{Corrections to scaling} 
To obtain a reliable estimate of  $\beta(t{\rightarrow}\infty)$ 
we analyze the transients, i.e. pre-asymptotic corrections. 
Our data suggests the following functional form:
\begin{equation}
  \label{e:dx_corr}
  \Delta x \approx a t^{2/3}\left(1 + bt^{-1/3}\right)\text{,}
\end{equation}
see Fig. \ref{fig:dx_corr}. By extrapolation of the numerical data (as indicated in the figure), we obtain $\Delta x \approx 1.125 \cdot t^{2/3}$ for the leading term. The numerical value of the prefactor will be discussed below. While the exponent of the subleading term in Eq. \eqref{e:dx_corr} is difficult to determine with certainty, an expansion of $\Delta x$ in powers of $t^{1/3}$ appears natural. 

\subsubsection{Spatial profile: Comparison with KPZ scaling}
\label{sec:spat-prof-comp}
We now turn to the analysis of how $\SF{x}{t}$ depends on the spatial coordinate $x$ and compare it with the relevant KPZ scaling function $\fkpz (x)$ \footnote{In the original context, \unexpanded{$\fkpz$} determines the asymptotic shape of the correlation function \unexpanded{$\langle v(x,t) v(x^{\prime},t^{\prime})\rangle$}, where \unexpanded{$v(x,t)$} denotes a solution of the stochastic Burger's equation and \unexpanded{$\langle \cdot \rangle$} averaging with respect to realizations of the noise.} Exact results for $\fkpz$ were obtained in Ref.\cite{prahofer_exact_2004} and its numerical values have been tabulated\cite{prahofer_exact_nodate}. $\fkpz$ resembles a Gaussian for small arguments, but it exhibits faster decay in the tails: 
$\fkpz(y) \sim \exp(- C {\times} |y|^3)$ for $|y| {\gg} 1$ with 
$C{\approx} 0.3$. 

\begin{figure*} 
\centering
  \mbox{    
\subfigure[\label{fig:resc_profile_log}]{\includegraphics[scale=0.4]{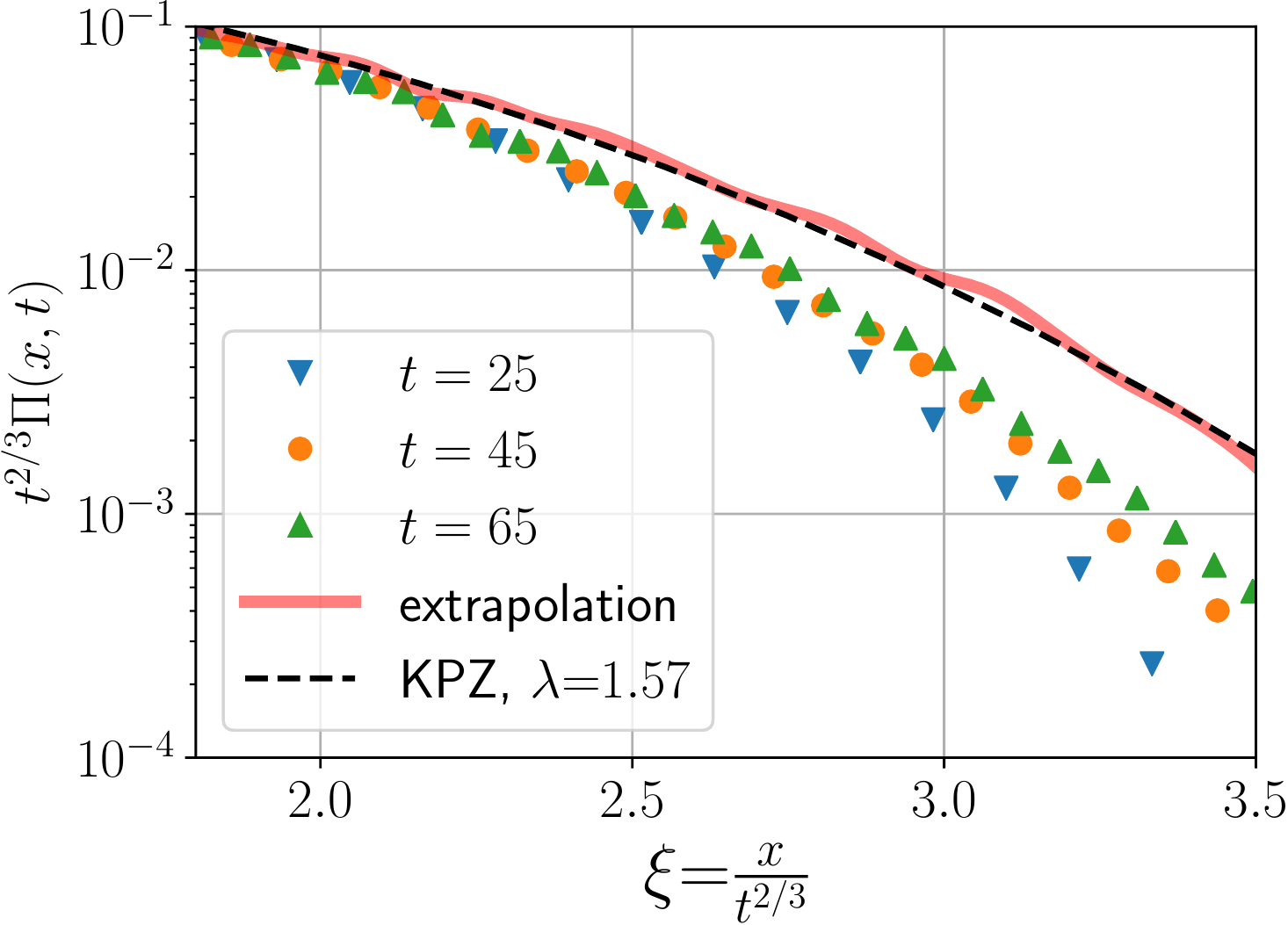}}    
\subfigure[\label{fig:resc_profile_lin}]{\includegraphics[scale=0.4]{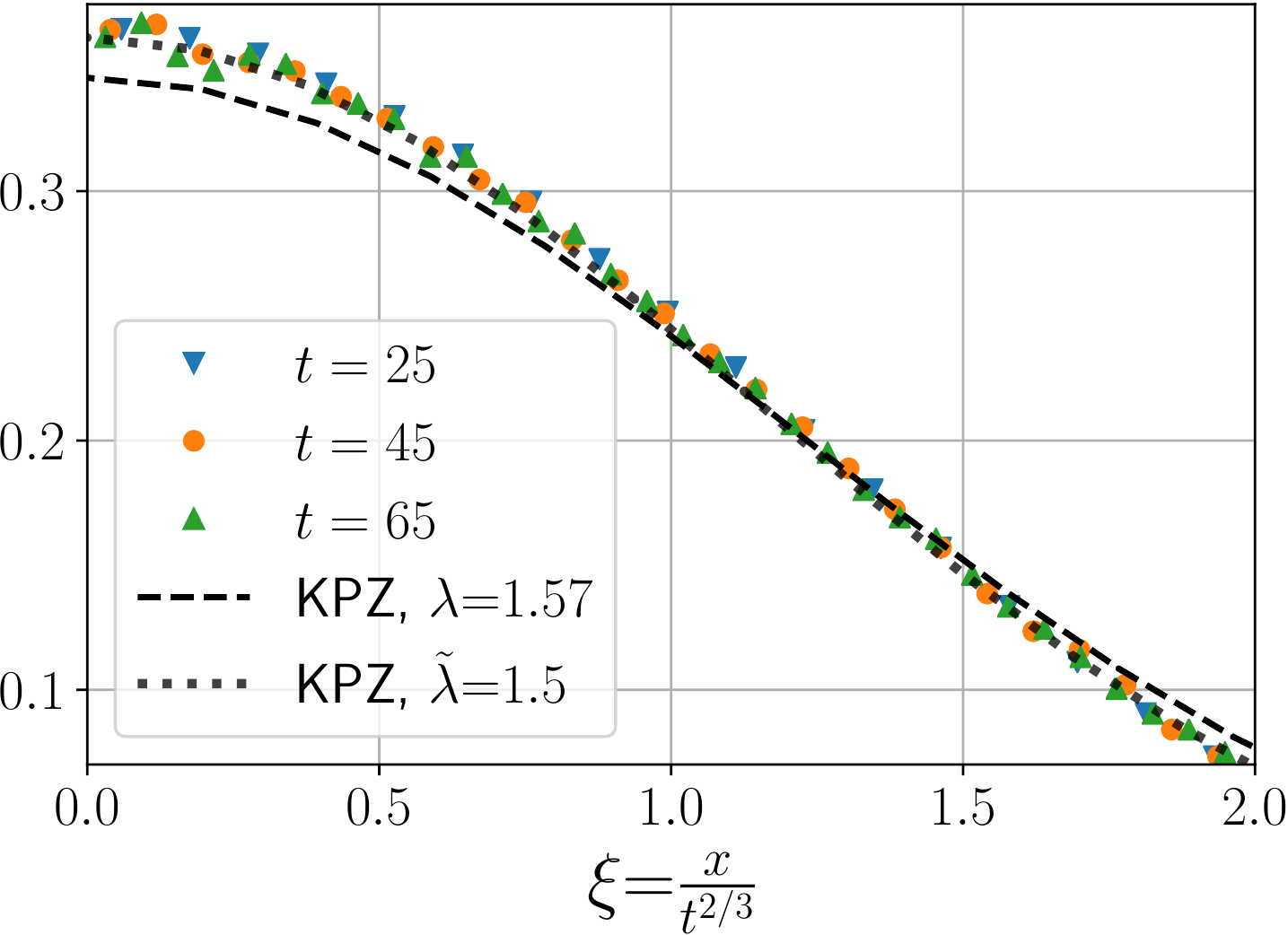}}
\subfigure[\label{fig:extrap}]{\includegraphics[scale=0.4]{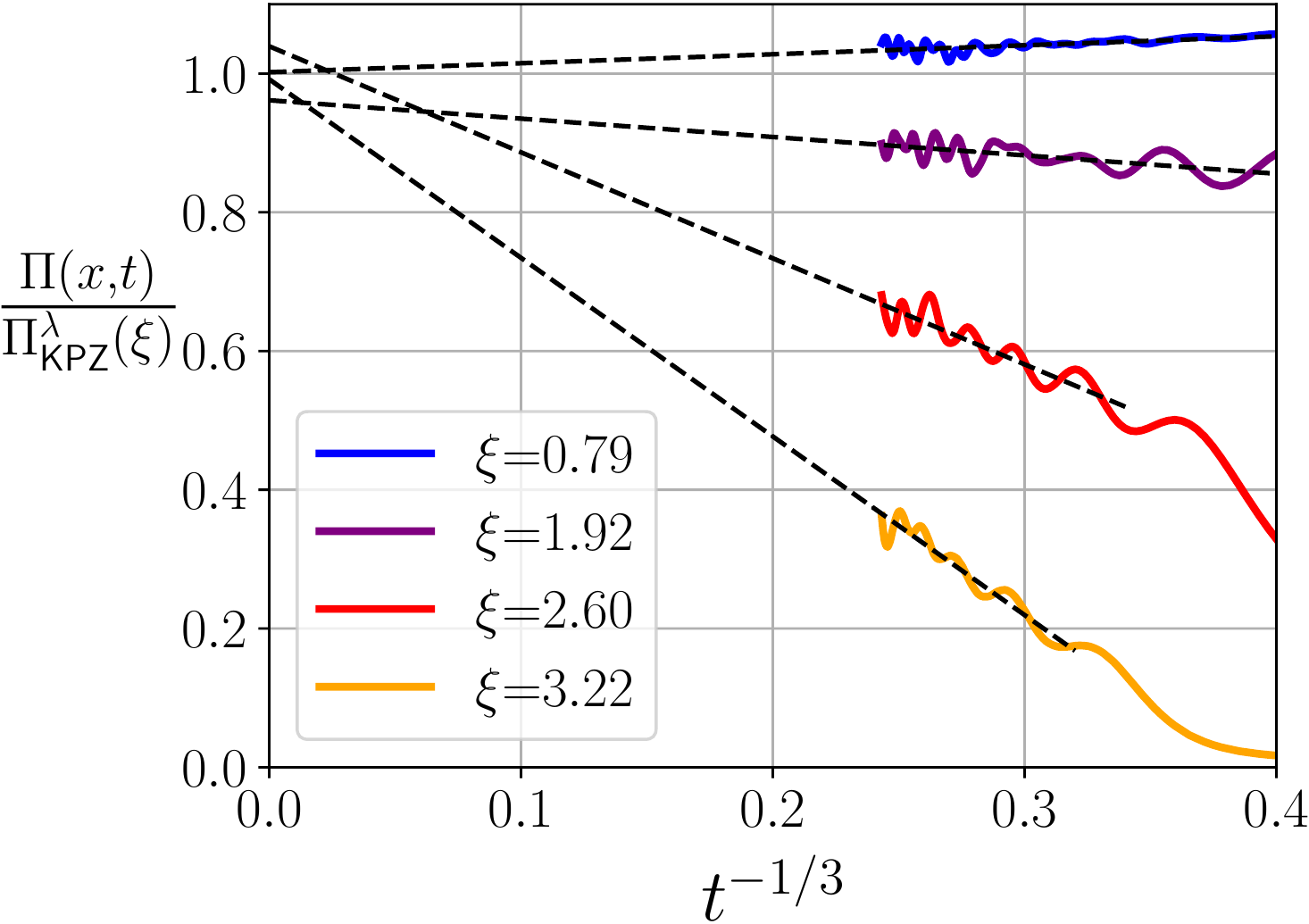}} 
}
\caption{\textbf{(a)} Numerical data for the tails of $\SF{x}{t}$, rescaled assuming a dynamical exponent of 
$z=\frac{3}{2}$, is compared to the KPZ prediction $\lambda^{-1}\fkpz(\frac{x}{\lambda t^{2/3}})$ taken from 
\cite{prahofer_exact_nodate}. Numerical results clearly exhibit a residual time dependence in the tails of $\Pi$. The 
parameter $\lambda=1.57$ has been chosen in order to be consistent with our extrapolation of $\Delta x(t)$. 
The important point of this figure is that the red line in (b), corresponding to an extrapolation with 
respect to time, agrees very well with the KPZ prediction. \textbf{(b)} The center of the correlation function $\Pi$ can 
be fitted by the KPZ scaling function with  $\lambda = 1.5$. The value of $\lambda$ obtained from such a fit is still 
time dependent on the time scales shown here (see discussion in the main text). \textbf{(c)} Illustration of the 
extrapolation scheme, which gives rise to the values of the red line shown in (a). Extrapolation is carried out for a 
fixed 
$\xi=\frac{x}{t^{2/3}}$ (after interpolating numerical data) presuming that corrections to scaling follow Eq. \eqref{e:corrections_pi}. The irregular oscillations in the data, which are observed at longer times, are a signature of truncation errors (see Appendix \ref{sec:convergence}). 
\label{f2}}
\end{figure*}

KPZ universality would imply the following scaling form of the spin correlator:
\begin{equation}
\label{e:scaling} 
\SF{x}{t} =  \frac{1}{\lambda t^{1/z}} \fkpz \left(\frac{x}{\lambda t^{1/z}}\right) \equiv \SFKPZ{\frac{x}{t^{1/z}}}
\end{equation}
with $z=\frac{3}{2}$. Our results for $\SF{x}{t}$ are shown in Fig. \ref{f2}
as a function of the scaling variable $\xi {=} \frac{x}{t^{2/3}}$.
Before discussing the the spatial dependence of the correlator, we recall our earlier result suggesting $\Delta x(t)/t^{2/3}\approx 1.125$ in the long-time limit (see Eq. \eqref{e:dx_corr}). Presuming KPZ scaling, we can relate the asymptotic time dependence of $\Delta x$ to the parameter $\lambda$ via
\begin{equation}
  \label{e:dx_lambda}
  \frac{\Delta x(t)}{t^{2/3}} \overset{t{\to}\infty}{\longrightarrow}\lambda \left( \int dy\,y^2\fkpz(y)\right)^{1/2} 
  \!\!\approx 0.715 \cdot \lambda \,\text{,}
\end{equation}
which implies $\lambda\approx 1.125/0.715\approx 1.57$. The corresponding prediction for the correlator $\Pi(x,t)$ together with the numerical data is shown in Fig. \ref{fig:resc_profile_lin}. We observe that the spatial shape based on KPZ scaling, $\Pi_{\text{KPZ}}^{\lambda=1.57}$, deviates from the numerical data for the relatively short times $t=35,45,65$ shown here. However, the results for $t^{2/3}\Pi(\xi)$ still exhibit a time dependence, which is most easily seen in the tails of the correlator (see Fig \ref{fig:resc_profile_log}). Consistent with the analysis above, we suggest that such finite time corrections vanish as $t^{-y}$ with $y=\frac{1}{3}$: 
\begin{equation}
  \label{e:corrections_pi}
  \SF{x}{t} = \SFKPZ{\xi}\left(1 + g(\xi)t^{-1/3}\right)\,\text{,}
\end{equation}
which indeed yields an accurate and consistent description of the numerical results, as can be seen from Fig. 
\ref{fig:extrap}. 
We note that, the numerical data for $\xi \lesssim 2 $, 
on the time scales shown in Fig. \ref{fig:resc_profile_lin}, 
agrees well with a KPZ scaling corresponding to an effective 
value $\tilde{\lambda}\approx 1.5$. At ever longer times this effective scale, 
$\tilde \lambda$, will eventually converge to $\lambda$ proper.
The small deviation seen in Fig. \ref{fig:resc_profile_lin} is not inconsistent with our analysis, 
but rather a trivial consequence of the normalization of $\Pi(x,t)$. 
\footnote{A similar time dependence of $\lambda$ has 
been observed in numerical works demonstrating KPZ scaling in classical models, see Refs. 
\cite{mendl_low_2015,kulkarni_fluctuating_2015}}.

Alternative interpretations based on a subleading power law with a different exponent $y>\frac{1}{3}$, as discussed in the Appendix \ref{sec:altern-scen-corr}), are possible. On the other hand, it seems unlikely that the corrections decay even slower than $t^{-1/3}$. Therefore, our estimate $\lambda=1.57$ could be considered an upper bound for the possible values of $\lambda$ that are still consistent with the numerics.    

The authors of Ref. \cite{ljubotina_kardar-parisi-zhang_2019} conjectured that $\lambda$ is exactly given by 
$\frac{3}{2}$, based on their numerical results. This conjecture appears inconsistent with our analysis. However, we 
suspect that the employed bond dimensions are not chosen sufficiently large in order to properly capture the transients. 
In fact, our results indicate that smaller bond dimensions tend to underestimate corrections to scaling (see Fig. 
\ref{fig:dx_analysis}). It is also interesting to note that Ref. \cite{das_kardar-parisi-zhang_2019} reports a very 
similar value of $\lambda \approx 1.55$ for the integrable classical analog of the \xxx chain at high temperatures.  


\subsection{Return probability\label{sec:fate-kpz-peak}}

The ``return probability", $\Pi(x{=}0,t)$, 
is a probe of the central peak. 
At zero magnetization $\sz {=} 0$, this peak exhibits the anomalous KPZ scaling. 
It is  interesting to inquire to what extent the anomalous scaling survives at 
finite $m$ and how the crossover, $m{\to}0$, occurs. This question has been addressed
in a recent work\cite{gopalakrishnan_anomalous_2019} by means of kinetic theory 
as well as MPO numerics. In the following, we present an analysis of our numerical 
data and thereby confirm some of the conclusions of Ref. \cite{gopalakrishnan_anomalous_2019}.

\begin{figure}[h]
  \centering
  \includegraphics[scale=0.52]{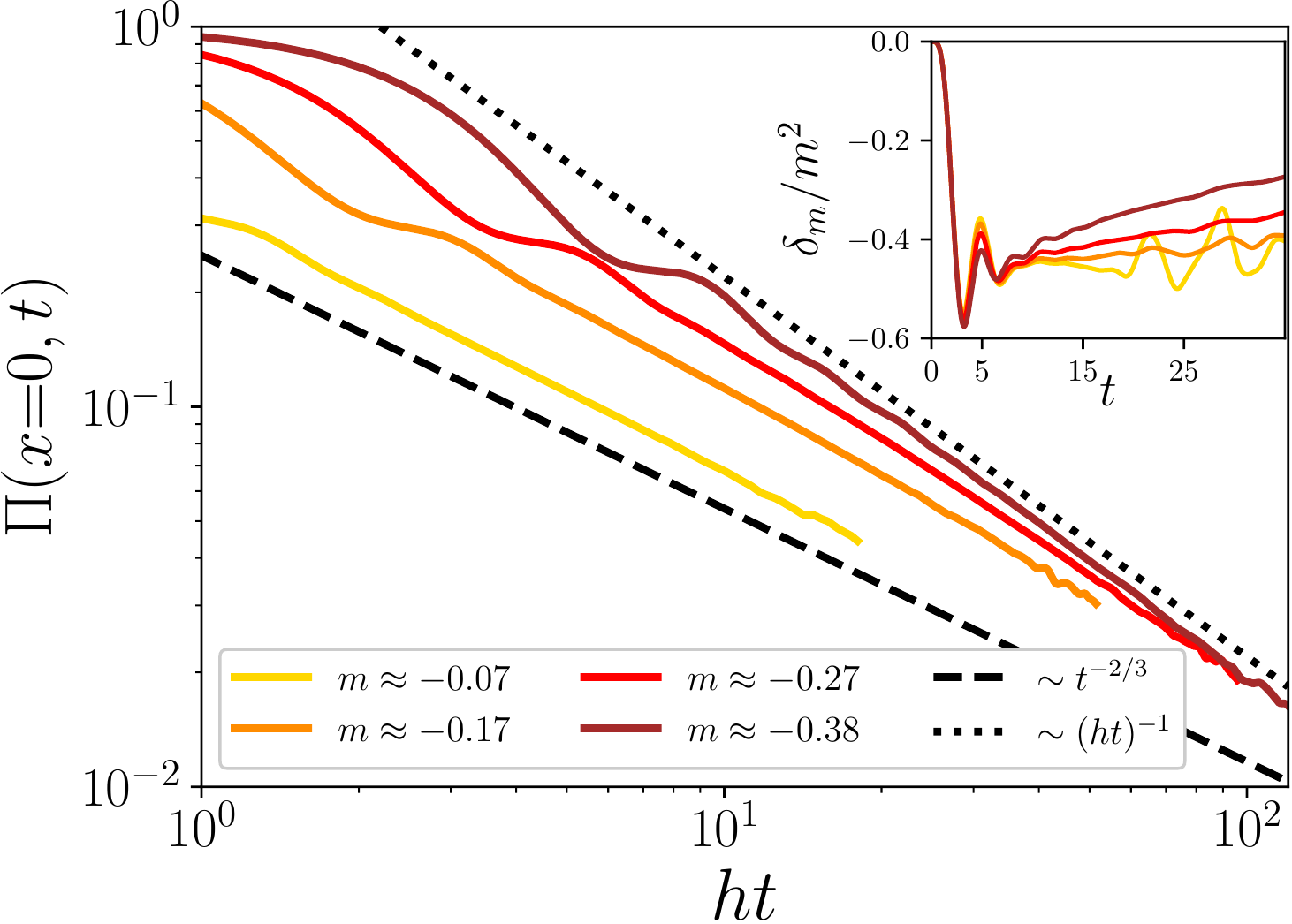}
\caption{
Double logarithmic plot of the return probability  
exhibiting the long-time, crossover behavior. 
The horizontal axis is rescaled in order to highlight the
 $\frac{1}{ht}$ behavior at long times. Black lines serve as guides to the eye. 
 \textit{Inset:} difference between the return probability $\Pi(x{=}0,t)$ at finite $m$ and $m=0$ scaled by $m^2$. The data (nearly) collapses at short times $t\lesssim 10$; 
at larger times a plateau develops for very small magnetizations $m(h)$. 
The fluctuations  (``noise'') seen in the small-$m$-data at larger times
are expected to disappear in the limit of large $\chi$ ($\chi=800$ was used here).
\label{fig:rp}
}
\end{figure}

\paragraph*{Short and intermediate times.}
As a measure for the impact of finite $m$, we define  
\be
\delta_m(t) = \Pi(0,t)\rvert_m -\Pi(0,t)\rvert_{m=0}. 
\ee
At $|m|$ relatively small, one expects a low impact only, as long as times are not too large, 
so assuming analyticity: $\delta_m(t) \propto m^2$. As shown in Fig. \ref{fig:rp} (inset), 
this is consistent with the simulation data in the window $0<t\lesssim 10$. 
The interpretation is straightforward: 
outmoving magnon modes carry spectral weight away from the center peak. 
At larger times and at $|m|$ small enough, we observe a 
plateau in $\delta_m(t)$, i.e. in this time window 
$\Pi(x=0,t)\rvert_m = \Pi(x=0,t)\rvert_{m=0} - C\times m^2$ with $C\approx 0.45$. 
These findings underline  that the anomalous KPZ-type behavior 
appears on an intermediate time scale once the magnetization $|m|$ is small enough, as one would expect.  


\paragraph*{Long times.}
The time dependence of $\Pi(x{=}0,t)$ at longer times is displayed 
in Fig. \ref{fig:rp}. 
The horizontal axis is rescaled in order to highlight the expected 
ballistic behavior,  
\begin{equation}
\Pi(x{=}0,t)\rvert_m \sim \frac{1}{ht},\,h=2\arctanh(2m)
\end{equation}
which reflects that magnonic quasi-particles carry away a finite magnetization at  $|m|>0$\cite{gopalakrishnan_anomalous_2019}. 
The crossover from KPZ to ballistic behavior is expected at times $t^{-2/3}\sim (ht)^{-1}$, so $t^{\ast}\sim h^{-3}$, consistent with \cite{gopalakrishnan_anomalous_2019}. 
At small values, $h\ll 1$, the time $t^{\ast}$ is well outside of our observation window. 
In this time window, the data displayed in Fig. \ref{fig:rp} is still consistent with an anomalous power law  $\Pi(x=0,t) {\sim} t^{-2/3}$. 
We do not actually observe a proper crossover in our numerical results, which would require long simulation times at a relatively small $|m|$ and, therefore, large bond dimensions $\chi$.
We interpret the curvature of the data shown in Fig. \ref{fig:rp} (on a doubly logarithmic scale), e.g. for $m{\approx}-0.17$, as an indication for the existence of such a crossover. 

\section{Summary and Outlook}
\label{sec:summary}

We have presented a comprehensive discussion of the $S^z$ autocorrelation function, 
$\Pi(x,t)$, for the spin-$\frac{1}{2}$ \xxx Heisenberg chain
at high temperature and fixed magnetization density, $m$.  
For any finite $|m| > 0$, the correlator $\Pi$ exhibits left- and right-moving peaks
that we attributed to magnon-type quasi-particles. The time dependence of the broadening associated with those peaks exhibits different behavior depending on $|m|$: Near maximum magnetization, $|m|{\lesssim}\frac{1}{2}$, the broadening follows a sub-diffusive $t^{1/3}$ scaling within our window of observation times, which we assign to
(cubic terms of) the bare quasi-particle dispersion. For weaker magnetization, $|m|\ll \frac{1}{2}$ a long time regime emerges with $t^{1/2}$-broadening that we loosely interpret as a signature of quasi-particle scattering. We interpret our results in terms of a transient behavior for $t<t_c$ following a $t^{1/3}$ law, which gives way to a $t^{1/2}$ law at long times $t>t_c$. Our results are consistent with a crossover time scale $t_c \sim (1/4-m^2)^{-3}$. 

At small $|m|$, a broad center peak dominates $\Pi(x,t)$.
The return probability, $\Pi(0,t)$, characterizes the corresponding dynamics. 
Also here, we find different behavior depending on $|m|$: ballistic decay, $t^{-1}$, is 
observed for the larger values of $|m|$, presumably reflecting the loss of amplitude due to the 
outgoing quasi-particles. On the other hand, close to $|m| = 0$, $\Pi(0,t)$ decays in a subballistic fashion following a $t^{-2/3}$ behavior. Our results are consistent with a 
crossover time $t^{\ast} \sim h^3$ between the two regimes, in agreement with the prediction of Ref. \cite{gopalakrishnan_anomalous_2019}. 

At zero magnetization, $m=0$, the propagating peaks are absent. 
The width of the correlator can be described by $\Delta x \approx a\cdot t^{2/3}(1 +b \cdot t^{-y})$,
with $a{\approx}1.125$, $b{\approx}-0.26$ and $y{=}\frac{1}{3}$. 
Motivated by the $t^{2/3}$-phenomenology, recent numerical work has tested $\Pi(x,t)$ 
against KPZ-scaling and indeed demonstrates matching with the KPZ scaling function\cite{ljubotina_kardar-parisi-zhang_2019}. 
We confirm this result after including finite time corrections. 
At this point, it seems that a deeper understanding of why 
KPZ scaling emerges in this model still needs to be developed
in future research. Such understanding appears even more relevant as recent works suggest 
that the KPZ behavior does not only occur in the spin-$\frac{1}{2}$ chain but in a large class
of integrable systems\cite{de_nardis_anomalous_2019,das_kardar-parisi-zhang_2019,dupont_universal_2019}.

\begin{acknowledgments}
The authors acknowledge valuable discussions with H. Spohn, A. L{\"a}uchli, B. N. Narozhny and thank T. Prosen for a helpful correspondence on computational details. PS acknowledges support by ERC-StG-Thomale-TOPOLECTRICS-336012. SB acknowledges support from SERB-DST, India, through Ramanujan
Fellowship Grant No. SB/S2/RJN-128/2016, Early Career Research Award
ECR/2018/000876 and MPG for funding through the Max Planck Partner Group at IITB. FE and FW acknowledge support by the DFG under Grants No.
EV30/11-1, EV30/12-1 and SFB-1277, project A03. We furthermore acknowledge
support by I. Kondov and computing time on the supercom-
puter ForHLR funded by the Ministry of Science, Research
and the Arts Baden-Württemberg and by the Federal Ministry
of Education and Research.
\end{acknowledgments}

\appendix

\begin{figure*}
  \centering
  \subfigure[\label{fig:prof_conv_lin}]{\includegraphics[scale=0.39]{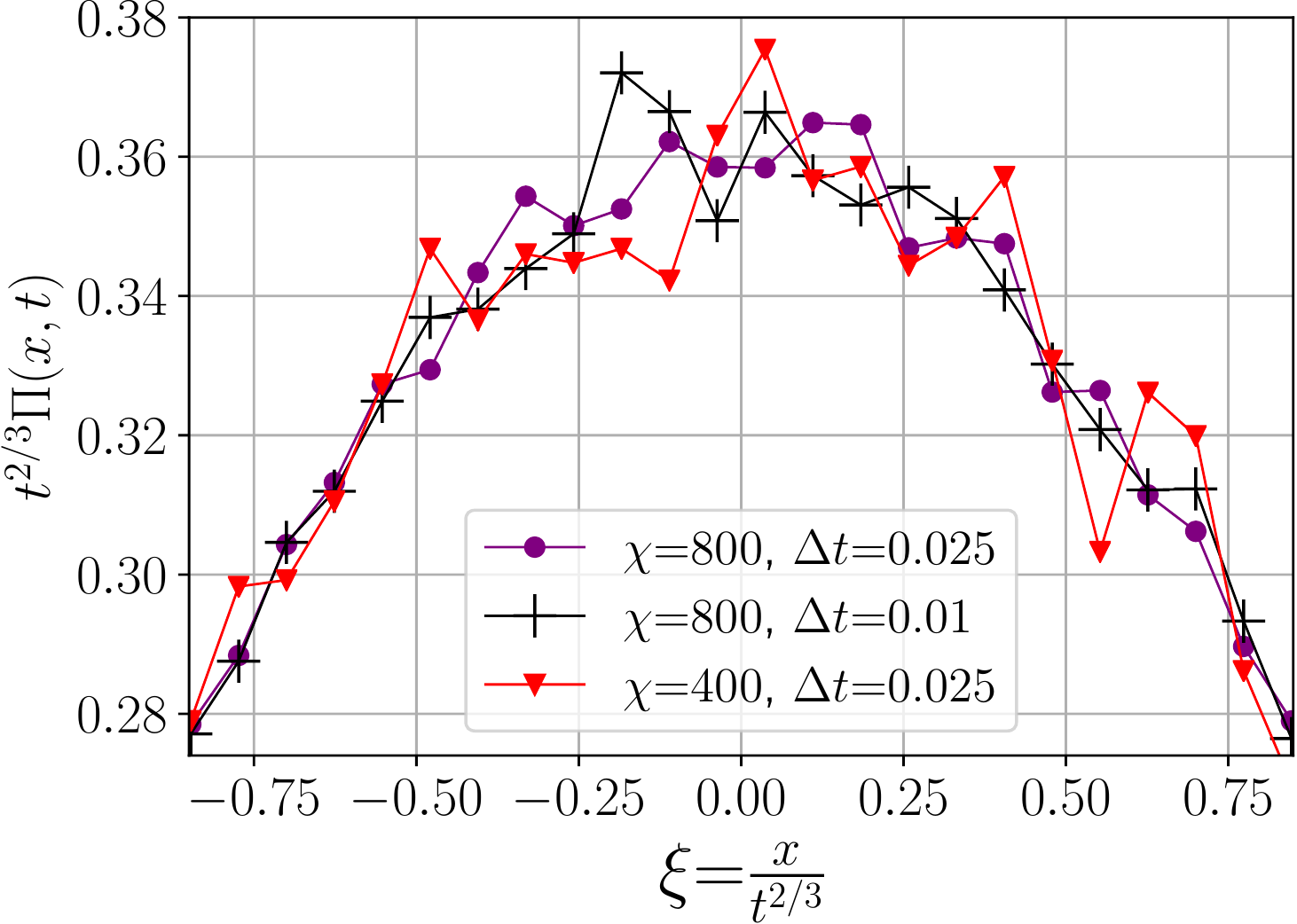}}
  \subfigure[\label{fig:prof_conv_log}]{\includegraphics[scale=0.39]{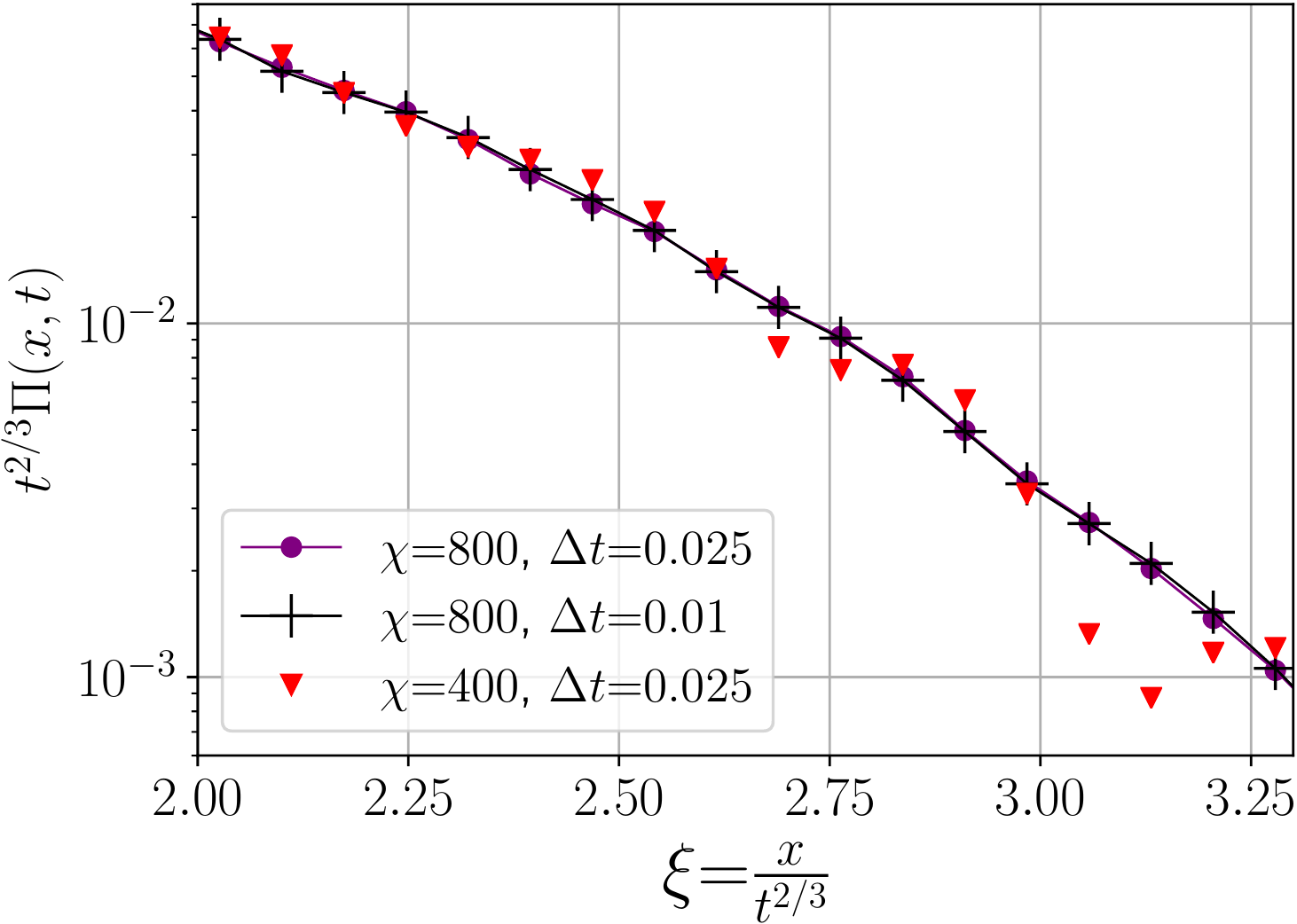}}
  \subfigure[\label{fig:beta_conv}]{\includegraphics[scale=0.39]{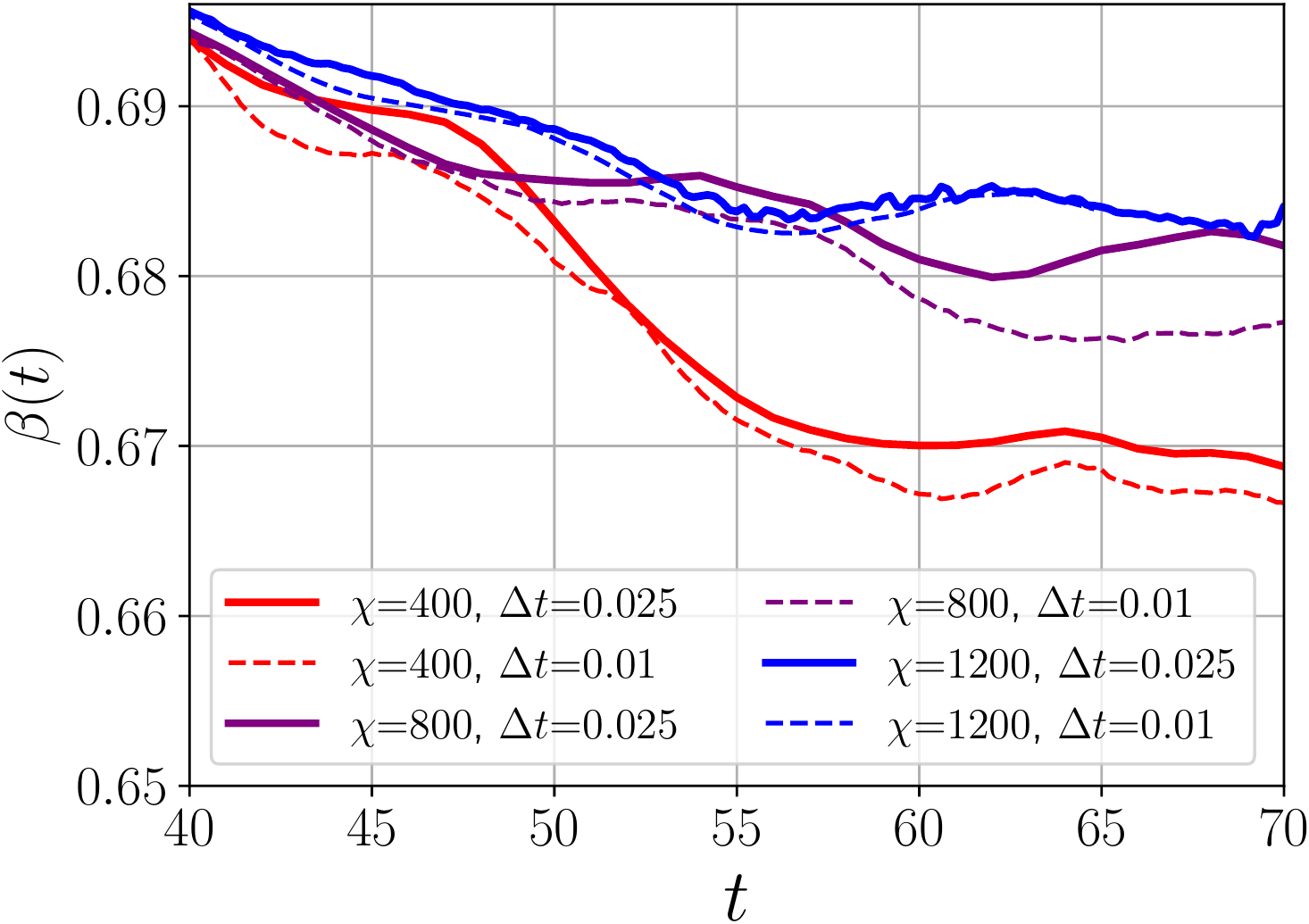}}
\caption{\textbf{(a),(b)} Rescaled correlator ($h=0$) at fixed time $t=50$ for various bond dimensions $\chi$ and Trotter time increments $\Delta t$. Truncation errors due to finite $\chi$ manifest themselves as unphysical spatial fluctuations, which are most pronounced near the center $x \sim 0$. At a given time, the tails are generally less affected by the truncation and are almost independent of $\Delta t$. \textbf{(c)} The effective exponent as shown in Fig. \ref{fig:beta} together with additional data obtained by using a smaller Trotter increment $\Delta t = 0.01$. Up to times $t\sim 60$, the data is not very sensitive to varying $\Delta t$. 
\label{fig:profile_conv}
}
\end{figure*}

\section{Matrix product state techniques}
\label{sec:matrix-product-state}

In this section, we briefly review the techniques employed to compute the time evolution of the initial state \eqref{e:init_state} and present an additional discussion of convergence properties. For details we refer to reviews of the topic, e.g. Refs. \cite{schollwock_density-matrix_2011,paeckel_time-evolution_2019}.
\subsection{Mixed state representation}
\label{sec:mixed-state-repr}

The matrix product representation of any operator (a so-called matrix product operator (MPO)) is equivalent to a matrix product state with an enlarged local Hilbert space\cite{zwolak_mixed-state_2004}. We choose the standard basis in operator space as a local basis set: $\Oket{0}=\dyad{\downarrow}{\downarrow},\,\Oket{1}=\dyad{\downarrow}{\uparrow},\,\Oket{2} = \dyad{\uparrow}{\downarrow},\,\Oket{3} = \dyad{\uparrow}{\uparrow}$. Then, a generic MPS representation (in operator space) of an operator $\hat{A}$ reads 
\begin{align}
\label{e:MPS_dop}
&\hat{A} \mathrel{\widehat{=}} \Oket{\hat{A}}= \sum_{\{\Sigma\}} A_1^{[\Sigma_1]}A_2^{[\Sigma_2]}\dotsm A_L^{[\Sigma_L]} \Oket{\{ \Sigma \} }, \\
&\Sigma_i \in \left\{0,1,2,3\right\} \nonumber
\end{align}
where $A_i^{[\Sigma]}$ denote matrices of dimensions $\chi_i \times \chi_{i+1}$, $\chi_i \leq \chi_{\text{max}}$ and $\chi_{1,L}=1$. $\chi_{\text{max}}$ denotes the maximum bond dimension of the MPS. In practice, we do not represent the density matrix in MPS form form but its square-root. This enforces positivity of the physical density operator and it allows to write the expectation values of observables in the same form as for pure states:
\begin{equation}
  \label{e:exp_val_sqroot}
   \Tr\left( \dop \hat{O} \right) = \Tr\left( \sqrt{\dop}\, \hat{O} \sqrt{\dop} \right) = \Obra{\sqrt{\dop}} \mathcal{O} \Oket{\sqrt{\dop}}
 \end{equation}
Here, $\mathcal{O}$ denotes a superoperator extension of $\hat{O}$ and the natural scalar product in operator space is given by the Frobenius product $\Obraket{\hat{A}}{\hat{B}} = \Tr\left(\hat{A}^{\dagger}\hat{B}\right)$. 

\subsection{Time evolution}
\label{sec:time-evolution}
The initial state \eqref{e:init_state} corresponds to a trivial MPO, i.e. it is a product state in operator space. As a close system is considered, its time evolution is governed by von-Neumann equation $\iu \partial_t \dop(t) = \comm{\hat{H}}{\dop} \mathrel{\widehat{=}} \mathcal{L}\Oket{\dop(t)}$, where $\mathcal{L}$ denotes the superoperator $\mathcal{L}\hat{O}=\comm{\hat{H}}{\hat{O}}$. Using this notation, we can introduce the analogue of the time evolution operator 
\begin{equation}
   \label{e:TEvoOp}
  \Oket{\dop(t)} = \mathcal{U}(t)\Oket{\dop(0)} = \exp\left(-\iu \mathcal{L} t\right) \Oket{\dop(0)}\,\text{.}
 \end{equation}
For models with nearest-neighbor terms only (as considered here), $\mathcal{L}$ can be written as 
\begin{equation}
   \label{e:L-sum}
   \mathcal{L} = \sum_{x=1}^{L-1} \mathcal{L}_{x,x+1}\,\text{,}
\end{equation}
where $\mathcal{L}_{x,x+1}$ acts on sites $x,x+1$ only. Then, $\mathcal{U}(\Delta t)$ can be subjected to a Suzuki-Trotter decomposition, e.g. of second order
 \begin{align}
   \label{e:ST-decomp}
   \begin{split}
     \mathcal{U}(\Delta t)= & e^{-\iu \mathcal{L}_{1,2} \Delta t/2}\dotsm e^{-\iu \mathcal{L}_{L-1,L} \Delta t/2}e^{-\iu \mathcal{L}_{L-1,L} \Delta t/2} \\
     & \dotsm e^{-\iu \mathcal{L}_{1,2} \Delta t/2} + O(\Delta t^3)
   \end{split}
 \end{align}
as used in this work. Truncation (in terms of singular values) is carried out simultaneously after each bond update in order to keep the bond dimensions below $\chi_{\text{max}}$ (simply denoted by $\chi$ throughout this work). The corresponding error is referred to as ``truncation error''. Throughout this work, we choose a very small cutoff for the singular values in the truncation procedure. Therefore, the maximum bond dimension alone controls the matrix product approximation in our simulations. 

\begin{figure}
  \centering
  \includegraphics[scale=0.53]{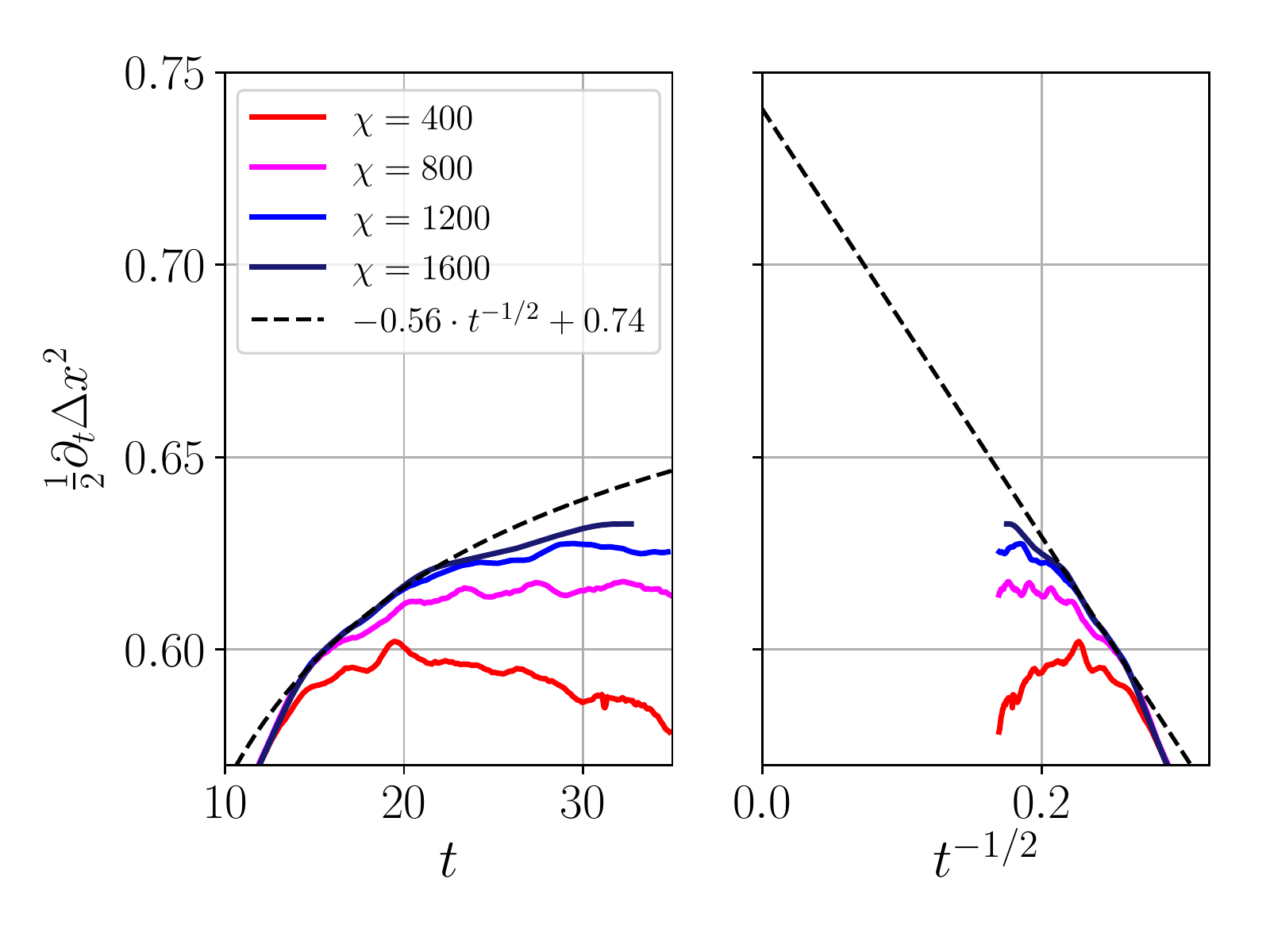}
  \caption{Temporal derivative of $\Delta x^2(t)$ for various values of the maximum bond dimension at anisotropy $\Delta{=}2$. The dashed line corresponds to an extrapolation assuming that $\partial_t \Delta x^2 = D + \text{const}\times t^{-1/2}$. The extrapolated value, $D \approx 0.74$, is consistent with the result of Ref.  \cite{de_nardis_diffusion_2019}.\label{fig:dc}}
\end{figure}

\subsection{Convergence\label{sec:convergence}}
In Fig. \ref{fig:profile_conv} we show additional data illustrating the dependence of the numerical results on $\chi$ as well as the Trotter time increment $\Delta t$. Choosing a smaller $\Delta t$ will decrease the error due to the Trotter decomposition of the time evolution operator. On the other hand, choosing a smaller value of $\Delta t$ requires a larger number of truncations to be carried out within a given window of time. Therefore, data obtained using a smaller $\Delta t$ is not necessarily more accurate. Furthermore, as soon as the results are not strictly converged with respect to $\chi$ (as is the case for the longest times shown in e.g. Fig. \ref{fig:beta}.), a dependence on the precise value of $\Delta t$ is also expected. However, we demonstrate in Fig. \ref{fig:profile_conv} that a certain degree of stability with respect to varying $\Delta t$ can be observed. 

In accordance with previous works \cite{ljubotina_kardar-parisi-zhang_2019,gopalakrishnan_anomalous_2019}, we observe that truncation errors generally introduce unphysical fluctuations in $\Pi(x,t)$. Those are most pronounced near $x=0$ while the fluctuations in the tails appear only at longer times in the form of more regular oscillations (see Fig. \ref{fig:prof_conv_log}). 

\begin{figure*}
  \centering      \subfigure[\label{fig:dx_corr2}]{\includegraphics[scale=0.39]{{dx_corrections}.pdf}\includegraphics[scale=0.39]{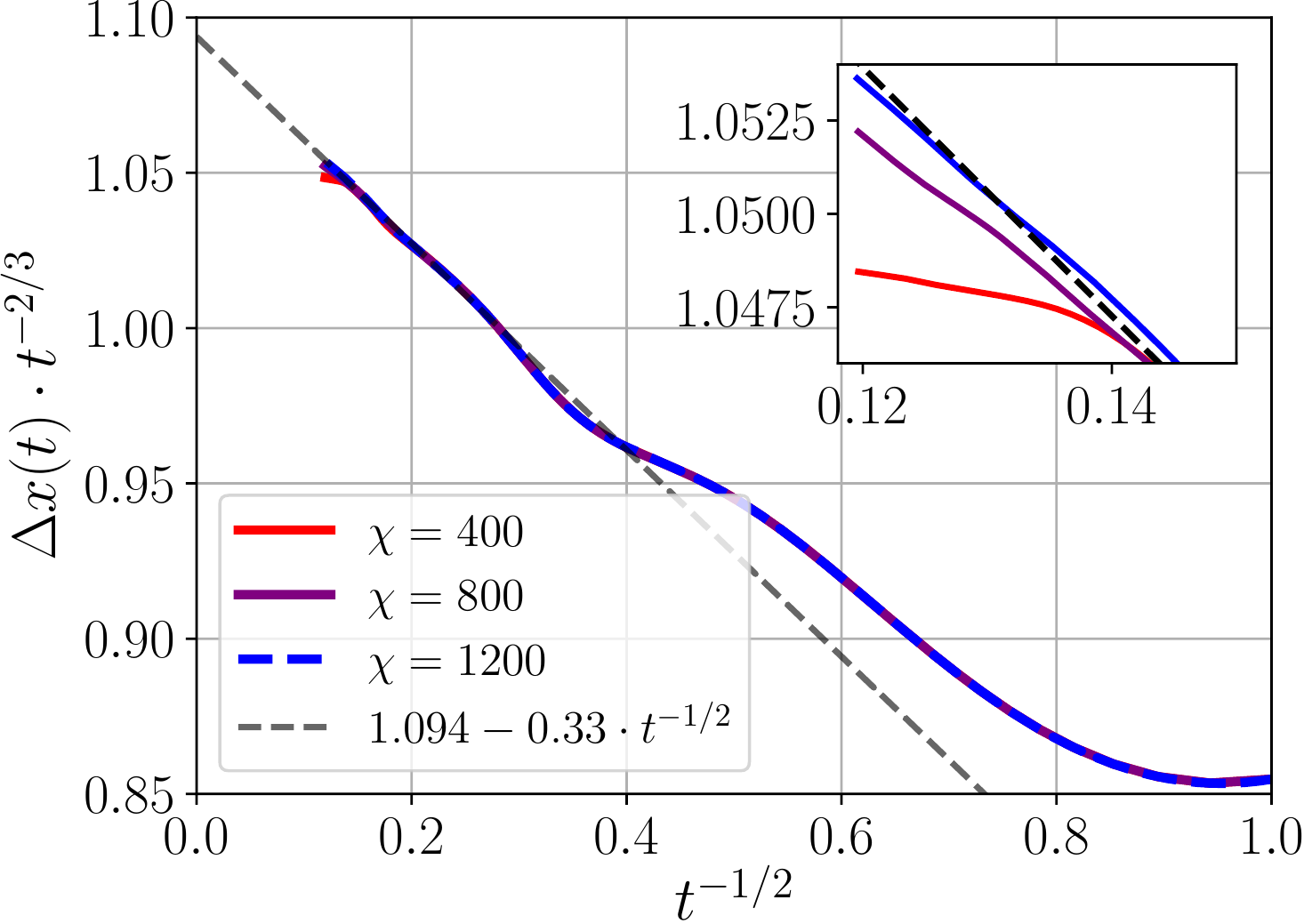}}
  \subfigure[\label{fig:prof_loglog_alt}]{\includegraphics[scale=0.39]{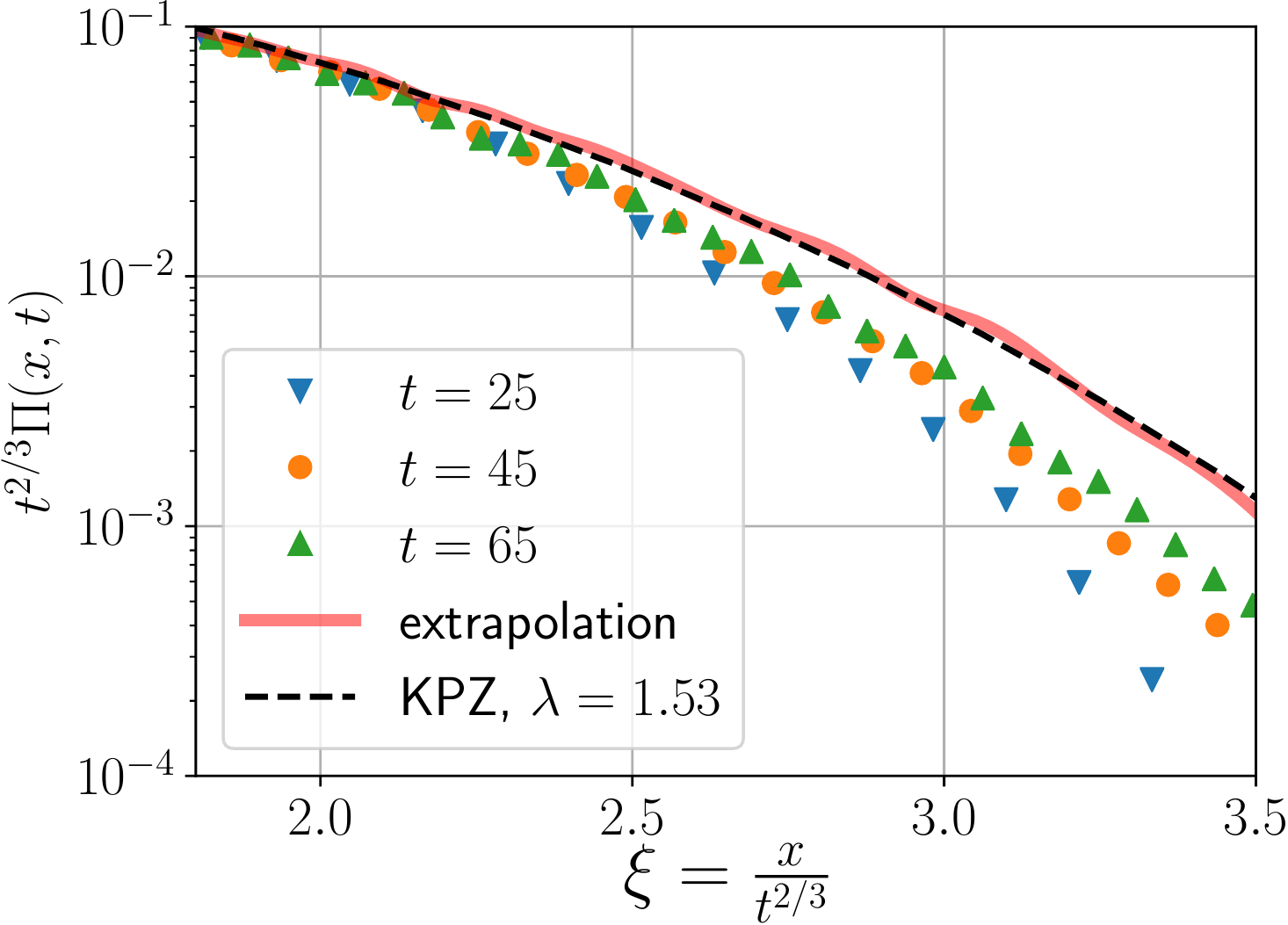}}
\caption{\textbf{(a)} Numerical results for $\Delta x(t)$ divided by the leading power law $t^{2/3}$. Different scenarios for the correction term are shown: $t^{-1/2}$ (left) and $t^{-1/3}$ (right, Fig. \ref{fig:dx_corr} is duplicated here for easier comparison). \textbf{(b)} Same data as in Fig. \ref{fig:resc_profile_log} with a different extrapolation (red line), which is consistent with the $t^{-1/2}$ correction.\label{fig:corrections_alt}}
\end{figure*}

\section{Diffusion constant at $\Delta = 2$ \label{sec:diff-const-at}}
For anisotropy $\Delta > 1$, the spin dynamics at vanishing magnetization $m=0$ is known to be normal diffusive \cite{de_nardis_diffusion_2019}. In the long-time limit, it is therefore expected that  $\partial_t\,\Delta x^2(t)\overset{t\rightarrow \infty}{=} 2Dt$ with $D$ denoting the diffusion constant. The results shown in Fig. \ref{fig:dc} demonstrate that this long-time limit cannot be reached reliably with bond dimensions $\chi < 2000$. A naive lower bound $D \gtrsim 0.63$ is obtained from this data by taking the maximum value reached for the largest bond dimension $\chi=1600$ available. This value should be contrasted with the value $D \approx 0.4$ shown in Ref. \cite{ljubotina_spin_2017} (cf. Fig. 2b, inset, of that reference), which employed the same protocol for simulating spin dynamics albeit with a much smaller bond dimension of $\chi = 200$. Our result is consistent with earlier works evaluating the diffusion constant by means of a direct evaluation of the current-current correlator at high 
temperature. In particular, for $\Delta = 2$, a lower bound of $D \gtrsim 0.56$ was given in Ref. \cite{karrasch_real-time_2014}, based on numerical data for $t \leq 17$ . Recently, Ref. \cite{de_nardis_diffusion_2019} obtained an analytic result of $D \approx 0.77$ and they give a numerical estimate of $D \approx 0.73$, which was obtained by an extrapolation scheme with respect to time. It is shown in Fig. \ref{fig:dc} that, applying the same extrapolation scheme, our data appears consistent with a very similar value of $D$. 

\section{Corrections to scaling: further discussion and alternative scenario}
\label{sec:altern-scen-corr}
In Section \ref{sec:spat-prof-comp}, we found that the numerical data for the spin correlation function can be described by $\SF{x}{t} = \SFKPZ{\xi}\left(1 + g(\xi)t^{-y}\right)$ with $y=\frac{1}{3}$ and $\lambda=1.57$. For comparison, we show an alternative scenario in Fig. \ref{fig:corrections_alt} corresponding to $y=\frac{1}{2}$ and $\lambda=1.53$, which also allows for consistent long-time extrapolations of $\Delta x(t)$ and $\Pi(x,t)$. However, the $t^{-1/3}$ correction appears to describe the time dependence of $\Delta x(t)$ more accurately down to very short times.

\end{document}